\documentclass[10pt,journal,doublecolumn]{IEEEtran}

\usepackage{graphicx}
\usepackage{algorithmicx,algorithm}
\usepackage{algpseudocode}

\usepackage{booktabs}

\usepackage[colorlinks,
linkcolor=black,
anchorcolor=black,
citecolor=black,
urlcolor = blue
]{hyperref}

\usepackage{amssymb}
\usepackage{subfigure}
\usepackage{multicol}
\usepackage{multirow}
\usepackage{amsmath}
\usepackage{bm}
\usepackage{cite}
\usepackage{gensymb}
\usepackage{amsfonts}
\usepackage{mathrsfs}
\usepackage{amsmath}
\usepackage{color}
\usepackage{balance}
\usepackage{bm}
\usepackage{setspace}

\usepackage{stfloats}
\usepackage{float}
\usepackage{epstopdf}

\usepackage{array}
\newcommand{\PreserveBackslash}[1]{\let\temp=\\#1\let\\=\temp}
\newcolumntype{C}[1]{>{\PreserveBackslash\centering}p{#1}}
\usepackage[flushleft]{threeparttable}
\usepackage{stfloats}
\usepackage{mathrsfs}
\usepackage{threeparttable}
\usepackage{setspace}

\newtheorem{lemma}{\bf Lemma}
\newtheorem{thm}{\bf Theorem}

\newtheorem{proposition}{\bf Proposition}

\DeclareMathOperator{\rank}{rank}

\begin{document}

\bibliographystyle{IEEEtran} 

\title{Channel Estimation for Optical Intelligent Reflecting Surface-Assisted VLC System: A Joint Space-Time Sampling Approach}

\author{Shiyuan Sun, Fang Yang, \emph{Senior Member, IEEE}, Weidong Mei, \emph{Member, IEEE},\\ Jian Song, \emph{Fellow, IEEE}, Zhu Han, \emph{Fellow, IEEE}, and Rui Zhang, \emph{Fellow, IEEE}

\begin{spacing}{0.3}
	\thanks{
		\scriptsize 
		
		S. Sun and F. Yang are with the Department of Electronic Engineering, Tsinghua University, Beijing 100084, P. R. China.
		(e-mail: sunsy20@mails.tsinghua.edu.cn; fangyang@tsinghua.edu.cn)
		\textit{(Corresponding author: Fang Yang)}

		W. Mei is the National Key Laboratory of Wireless Communications, University of Electronic Science and Technology of China, Chengdu 611731, China.
		(e-mail: wmei@uestc.edu.cn)

		J. Song is with the Department of Electronic Engineering, Tsinghua University, Beijing 100084, P. R. China, and also with the Shenzhen International Graduate School, Tsinghua University, Shenzhen 518055, P. R. China.
		(e-mail: jsong@tsinghua.edu.cn)
		
		Z. Han is with the Department of Electrical and Computer Engineering at the University of Houston, Houston, TX 77004 USA, and also with the Department of Computer Science and Engineering, Kyung Hee University, Seoul, South Korea, 446-701.
		(e-mail: hanzhu22@gmail.com)
		
		R. Zhang is with School of Science and Engineering, Shenzhen Research Institute of Big Data, The Chinese University of Hong Kong, Shenzhen, Guangdong 518172, China (e-mail: rzhang@cuhk.edu.cn). 
		He is also with the Department of Electrical and Computer Engineering, National University of Singapore, Singapore 117583. 
		(e-mail: elezhang@nus.edu.sg)
	}
\end{spacing}
}
\newenvironment{thisnote}{\par\color{blue}}{\par}

\maketitle
\begin{abstract}
Optical intelligent reflecting surface (OIRS) has attracted increasing attention due to its capability of overcoming signal blockages in visible light communication (VLC), an emerging technology for the next-generation advanced transceivers. 
However, current works on OIRS predominantly assume known channel state information (CSI), which are essential to practical OIRS configuration. 
To bridge such a gap, this paper proposes a new and customized channel estimation protocol for OIRSs under the alignment-based channel model. 
Specifically, we first unveil OIRS spatial and temporal coherence characteristics and derive the coherence distance and the coherence time in closed-form. 
Next, to achieve fast beam alignment over different coherence time, we propose to dynamically tune the rotational angles of the OIRS reflecting elements following a geometric optics-based non-uniform codebook. 
Given the above beam alignment, we propose an efficient joint space-time sampling-based algorithm to estimate the OIRS channel. 
In particular, we divide the OIRS into multiple subarrays based on the coherence distance and sequentially estimate their associated CSI, followed by a spacetime interpolation to retrieve full CSI for other non-aligned transceiver antennas. 
Numerical results validate our theoretical analyses and demonstrate the efficacy of our proposed OIRS channel estimation scheme as compared to other benchmark schemes.
\end{abstract}

\begin{IEEEkeywords}
Visible light communication (VLC), optical intelligent reflecting surface (OIRS), channel estimation, channel coherence, codebook design.
\end{IEEEkeywords}

\IEEEpeerreviewmaketitle
\begingroup
\allowdisplaybreaks

\section{Introduction}
\IEEEPARstart{W}{ith} successful industrialization of millimeter-wave (mmWave) communication in 5G, researchers continue to explore communication technologies in Terahertz (THz) and even higher frequency bands to address the scarcity of available spectrum~\cite{sarieddeen2020next}. 
Especially, visible light communication (VLC) that boasts nearly 400 THz frequency-band is regarded as a pivotal technology for 6G~\cite{chi2020visible}. 
VLC offers several advantages over current radio frequency (RF) communications, including its license-free spectrum, high energy efficiency, environmental friendliness, and ubiquitous next-generation advanced transceivers (NGAT) such as light-emitting diode (LED) and photodetector (PD)~\cite{burchardt2014vlc,karunatilaka2015led}. 
However, due to the nanoscale wavelength of lightwave, VLC faces significant challenges in signal coverage. The prior studies have shown that the penetration loss of lightwaves is almost infinite and communications behind the obstacles can hardly maintain due to diffuse reflections with negligible reflected energy~\cite{feng2018mmwave}. 

Fortunately, the advent of digital controllable meta-material, namely intelligent reflecting surface (IRS), offers a viable solution to deal with the blockage issue~\cite{gongshimin2020}. 
Specifically, an IRS consists of a multitude of reflecting elements that can alter the phase shift and/or strength of the incident wave by adjusting its load resistance or impedance~\cite{basar2019wireless,wu2021intelligent}. 
Thanks to such a capability, IRS can help eliminate the signal dead zones by bypassing the blockages in the environment~\cite{wu2019towards,9771079}. 
Nonetheless, IRS configuration heavily relies on the acquisition of channel state information (CSI), which is practically challenging due to the passivity of IRS, the high dimentional cascaded IRS channel, and the complicated IRS reflected channel model~\cite{zheng2022survey}. 
Consequently, developing an efficient IRS channel estimation method, especially in visible light frequency band, is an open problem to be solved.

\subsection{Prior Works}
Generally, the IRS channel estimation approaches in the RF communication can be classified into two categories~\cite{zheng2022survey}. 
\textit{(i) Cascaded channel estimation:} This approach considers fully passive IRSs without the ability to transmit/receive pilots, aiming to estimate the cascaded channel from the transmitter to the receiver via the IRS at the user side (for downlink) or at the base station side (for uplink). 
The cascaded channel estimation relies on the design of the pilot sequence and/or the training reflection pattern~\cite{mishra2019channel,zheng2020intelligent,jensen2020optimal}. 
Due to the cascaded nature of the channel, its dimension can become considerably high if the number of IRS reflecting elements and/or transmit/receive antennas is large. 
As such, some existing works have also explored how to leverage the sparsity of the high-frequency channels to reduce the estimation overhead~\cite{he2019cascaded,wang2020compressed,he2021channel}. 
\textit{(ii) Separate channel estimation:} In this approach, the IRS is equipped with several dedicated sensing components, which endow it with some basic signal processing capacbility, thus referred to as ``semi-passive IRS''~\cite{zheng2022survey}. 
Therefore, the channel estimation can be completed at the IRS side by employing some classical channel estimation techniques, such as the least-square or minimum mean square error (MMSE) estimators~\cite{hu2021semi}.

However, the above channel estimation approaches designed for RF IRSs cannot be directly applied to optical IRSs (OIRSs) due to the following reasons. 
First, the Lambertian wave model is adopted in VLC, as opposed to the planar wave model and the spherical wave model adopted in far- and near-field RF scenarios, respectively~\cite{obeed2019optimizing,zhang20236g}. 
Besides, unlike the subwavelength elements used for RF IRS~\cite{mi2022towards}, the OIRS reflecting elements are much larger than the lightwave wavelength; thus, the specular reflector approximation should be applied rather than the diffuse approximation as in RF IRSs~\cite{2021intelligent}.
Last but not least, the OIRS-reflected channel model differs significantly from the RF IRS-reflected channel model~\cite{abdelhady2020visible}. 
As such, dedicated OIRS channel estimation approaches should be developed, which, however, remains an open problem.

To date, two primary channel models have been proposed for OIRS-assisted VLC systems, namely, the physical optics model and the alignment-based channel model. 
Specifically, the former model directly characterizes the precise reflection channel gain in terms of the parameters of each reflecting element, i.e., the rotation angle of intelligent mirror array (IMA) or the phase discontinuity of the intelligent metasurface reflector (IMR)~\cite{abdelhady2020visible}. 
Focusing on this model, several studies have delved into the performance optimization problem for OIRS-aided wireless communications pertaining to single-user system~\cite{9543660}, physical layer security~\cite{qian2021secure}, nonorthogonal multiple access (NOMA)~\cite{9838853}, among others~\cite{aboagye2023ris,10183987,wu2022capacity}. 
Nonetheless, this model is hindered by its reliance on the complicated and nonlinear Lambertian-like expression~\cite{9543660}, which makes channel estimation difficult in practice. 
Alternatively, the alignment-based channel model is a more tractable approach~\cite{sun_CL,10190313}. 
Owing to the focused energy pattern of the OIRS-reflected signal~\cite{aboagye2023ris}, it is feasible to align each OIRS reflecting element with a specific pair of transceiver antennas by tuning its rotation angles (i.e., yaw and roll angles), referred to as the ``angle-selective'' property. 
As such, the OIRS-reflected channel can be characterized in terms of the associations/alignment between the OIRS reflecting elements and the transceiver antennas. 
Note that the channel parameters for this model are independent to the Lambertian formula and thus can be applied to the general pilot-based channel estimation. 
However, to the best of our knowledge, there is no existing work studying the OIRS channel estimation under any of the above channel models.

\subsection{Our Contributions}
To bridge the aforementioned gap, this paper proposes new OIRS channel estimation approaches for VLC system under the alignment-based channel model. 
The detailed contributions are summarized as follows:

\begin{itemize}
	\item
	We propose a new channel estimation protocol for the OIRS-assisted VLC system, including the dynamic beam alignment and sequential subarray channel estimations based on the OIRS coherence properties. 
	In particular, we exploit the intrinsic coherence in VLC channels due to the intensity modulation and direct detection (IM/DD) scheme and unveil the specific coherence characteristics of the OIRS-reflected channel in both spatial and temporal domains, based on which its coherence distance in terms of geometry and coherence time in presence of user mobility are derived in closed-form.

	\item
	To achieve fast beam alignment over different channel coherence time, we leverage the angle-selective property of each OIRS reflecting element by dynamically tuning its rotation angles to align with the specific transceiver antennas following a geometric optics-based OIRS codebook, where the angle codewords, i.e., the yaw and roll angle of each OIRS reflecting elements, are sampled in a non-uniform manner. 
	It is analytically shown that the proposed sampling method can significantly reduce the feasible codebook space with negligible performance loss and enable faster beam alignment than its uniform counterpart.

	\item
	Next, to efficiently estimate the OIRS-reflected channel for the aligned transceiver antennas given in each channel coherence time, we propose a joint space-time samplingbased (JSTS) algorithm, where all OIRS reflecting elements are divided into multiple subarrays based on the coherence distance to estimate their associated channels sequentially to avoid the high overhead for element-wise channel estimation. 
	Finally, a multi-dimensional interpolation is conducted to retrieve the full CSI including that for other non-aligned transceiver antennas. 
	Numerical results validate the coherence characteristics of OIRS-reflected channels and show the superiority of our proposed channel estimation protocol to other benchmarks.
\end{itemize}

The remainder of the paper is organized as follows. 
Section~\ref{Sec:Model} presents the OIRS-reflected channel model and our proposed OIRS channel estimation protocol. 
Section~\ref{Sec:Coherence} analyzes the OIRS coherence properties in spatial and temporal domains. 
Section~\ref{Sec:Codebook} presents the geometric optics-based OIRS codebook for fast beam alignment. 
Section~\ref{Sec:Proposed} presents our proposed JSTS algorithm for low-complexity OIRS channel estimation. 
Numerical results are presented in Section~\ref{Sec:numerical}, followed by conclusions drawn in Section~\ref{Sec:Conclude}.

\textit{Notations:}
Scalars are represented by italicized letters, such as $A$, while vectors and matrices are represented by lowercase and uppercase boldface letters, respectively, such as $\boldsymbol{a}$ and $\boldsymbol{A}$.
The uppercase boldface letters $\textbf{A} = (\textbf{A}_x,\ \textbf{A}_y,\  \textbf{A}_z)$ denote coordinates in a three-dimensional Cartesian coordinate system, with $\textbf{A}_x/\textbf{A}_y/\textbf{A}_z$ representing the $X$-/$Y$-/$Z$-axis component of $\textbf{A}$.
$\textbf{AB}$ denotes a vector from $\textbf{A}$ to $\textbf{B}$ and $\widehat{\textbf{AB}}$ denotes its normalized version with unit modulus, i.e., $\|\widehat{\textbf{AB}}\|_2=1$.
The matrix transpose, vectorization, inversion, Hadamard product, Kronecker product, ceiling function, and indicator function are denoted by $(\cdot)^T$, $\text{vec}(\cdot)$, $(\cdot)^{-1}$, $\odot$, $\otimes$, $\lceil \cdot \rceil$, and $\mathbb{I}(\cdot)$, respectively. 
The entry of a matrix $\boldsymbol{A}$ located in the $i$th row and the $j$th column is denoted as $[\boldsymbol{A}]_{i,j}$.
We use calligraphic letters, such as $\mathcal{A}$, to denote sets. 

\section{System Model}
\label{Sec:Model}
Following physical optics, this section first presents the channel gain of a single OIRS reflecting element (patch) and shows its angle-selective property. 
Then, the multi-patch OIRS-reflected channel presented, followed by the OIRS channel estimation protocol proposed.

\begin{figure}[t]
	\centering  
	\subfigure[Single-patch channel]{   
		\begin{minipage}[b]{0.37\textwidth} 
			\centering    
			\includegraphics[width=0.95\textwidth]{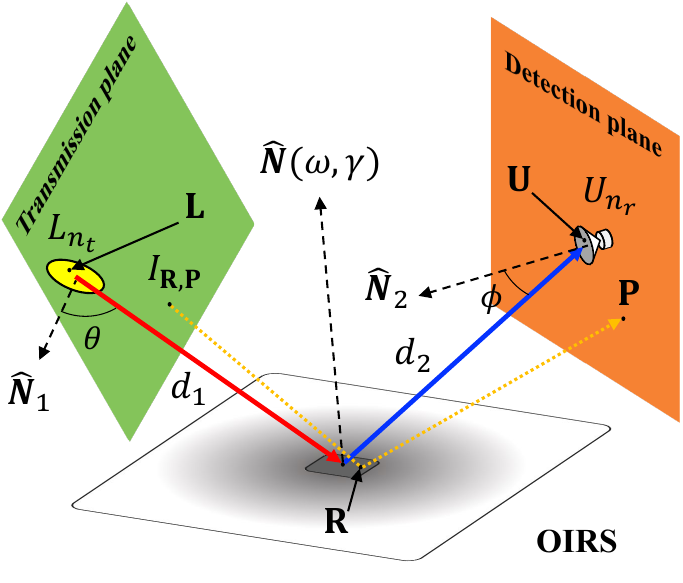}
			\label{Fig:single}
		\end{minipage}
	}
	\subfigure[Multi-patch channel]{ 
		\begin{minipage}[b]{0.37\textwidth} 
			\centering    
			\includegraphics[width=0.95\textwidth]{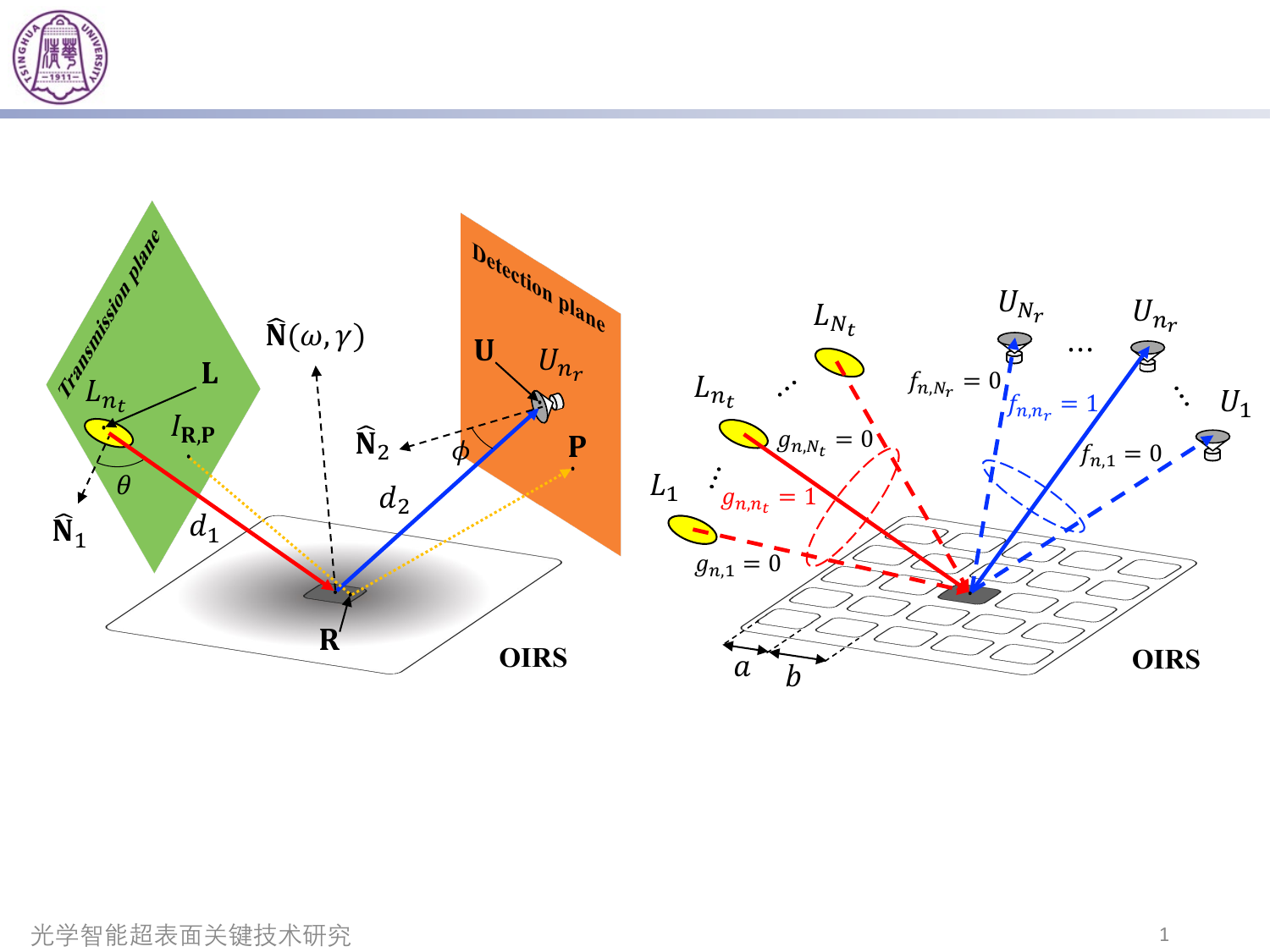}
			\label{Fig:multiple}
		\end{minipage}
	}
	\caption{The channel model of the OIRS-reflected VLC path.}    
	\label{Fig:channel}    
\end{figure}

\subsection{Single-patch Channel}
\label{Sec:Model single}

Generally, OIRS hardware architecture can be implemented by an array of mirrors in IMA or the optical meta-material in IMR. 
In this paper, we adopt the IMA due to its simplicity.
The physical parameters of an IMA can be represented by the roll angle $\omega \in [-\pi/2, \pi/2)$ and yaw angle $\gamma \in [-\pi/2, \pi/2)$ of each OIRS reflecting element.
As shown in Fig.~\ref{Fig:single}, consider a signal path that emits from an LED, then is reflected by a single OIRS reflecting element, and finally reaches a PD.
Let $\mathcal{L}$, $\mathcal{R}$, and $\mathcal{U}$ denote the point sets within the LED, the OIRS reflecting element, and the PD, respectively, with $|\mathcal{L}|$, $|\mathcal{R}|$, and $|\mathcal{U}|$ representing the corresponding areas.
Let $\textbf{L}\in \mathcal{L}$ and $\textbf{U} \in \mathcal{U}$ represent the center points of the LED and the PD, respectively.
Moreover, $\textbf{R}\in \mathcal{R}$ denotes an arbitrary point within the reflecting element and $\textbf{P}$ denotes an arbitrary point in the detection plane.
The normal vector of the OIRS reflecting element, which is denoted by $\widehat{\boldsymbol{N}}(\omega, \gamma)$, can be expressed as~\cite{abdelhady2020visible}
\begin{equation}
	\widehat{\boldsymbol{N}}(\omega, \gamma) = (\cos\omega \sin\gamma,\ \cos\omega \cos\gamma,\ -\sin\omega).
\end{equation}

Let the normal vectors of the LED and the PD be denoted by $\widehat{\boldsymbol{N}}_1$ and $\widehat{\boldsymbol{N}}_2$, respectively.
Let $I_{\textbf{R},\textbf{P}}$ denote their corresponding point in the transmission plane so that $I_{\textbf{R},\textbf{P}}$-$\textbf{R}$-$\textbf{P}$ forms a path following the reflection law.
According to~\cite[Eq. (52)]{abdelhady2020visible}, the power density at $\textbf{P}$ can be derived by accumulating the contributions of differential elements within $\mathcal{R}$ as
\begin{align}
	\setlength\abovedisplayskip{3pt}
	\label{Eq:single powerDensity}
	E(\textbf{P}; \omega, \gamma) = & E_0 \iint_{\textbf{R} \in \mathcal{R}}  \frac{\delta(m+1)\cos^{m-1}\theta}{2\pi |\mathcal{L}| \|\textbf{PR}\|_2^2} \widehat{\boldsymbol{N}}_2^T\widehat{\textbf{PR}} \notag\\
	& \times \mathbb{I}(I_{\textbf{R},\textbf{P}} \in \mathcal{L}) \times (-\widehat{\boldsymbol{N}}(\omega, \gamma)^T\widehat{\textbf{PR}}) \,d\textbf{R}_x\,d\textbf{R}_y,
	\setlength\belowdisplayskip{3pt}
\end{align}
\vspace{-0.2cm}

\noindent
where $E_0$, $\theta = \arccos(\widehat{\boldsymbol{N}}_1^T\widehat{\textbf{LR}})$, $\delta$, and $m$ denote the emitted power, the angle of irradiance, the reflectivity of the OIRS, and the Lambertian index, respectively.

\begin{figure}[t]
	\centering
	\includegraphics[width=0.44\textwidth]{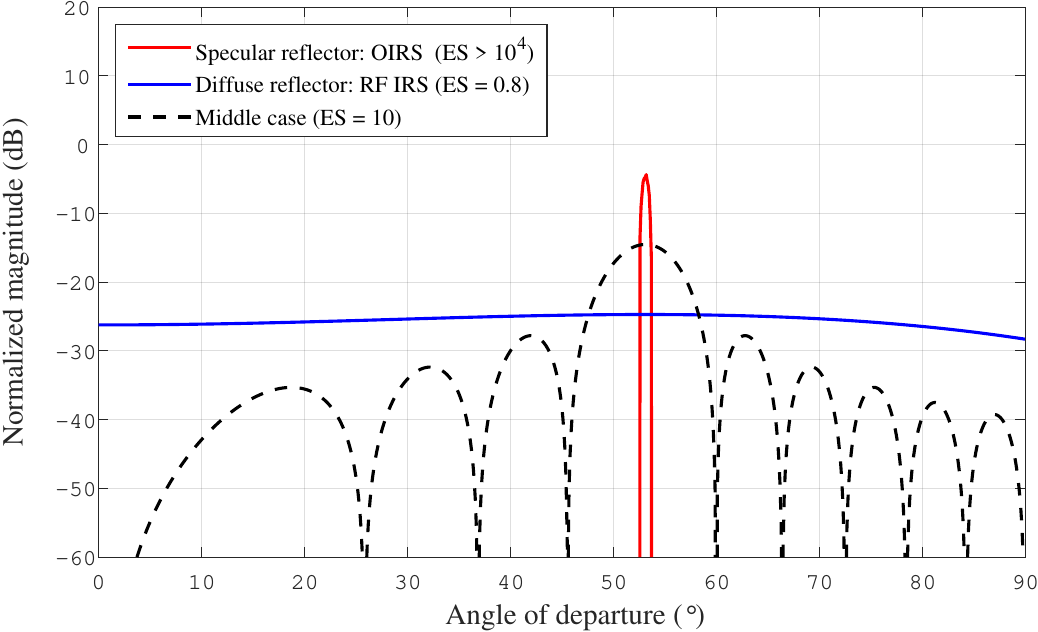}
	\caption{Normalized magnitude curves versus AoD for RF IRS and OIRS.}
	\label{Fig:electrical size}
	\vspace{-0.6cm}
\end{figure}

\footnotetext[1]{The ES is defined as the ratio of the metallic aperture to the signal wavelength.}

Based on~(\ref{Eq:single powerDensity}), numerical simulation is conducted to illustrate the differences between the OIRS- and RF IRS-reflected channel.
Specifically, we consider a circle LED with a radius of 5 cm centered at $\textbf{L}$ = (2m, 2m, 3m), a square IRS reflecting element with a length of 2 cm centered at $\textbf{R}$ = (2m, 0m, 1.5m), and a square PD with a length of 2 cm centered at $\textbf{U}$ = (2m, 2m, 0m).
All normal vectors $\widehat{\boldsymbol{N}}(\omega, \gamma)$, $\widehat{\boldsymbol{N}}_1$, and $\widehat{\boldsymbol{N}}_2$ are perpendicular to their respective planes.
The power density of the OIRS-reflected lightwave signal is calculated based on~(\ref{Eq:single powerDensity}), while that of the RF IRS-reflected signal is calculated based on~\cite[Eq. (4)]{ozdogan2019intelligent}, assuming the electrical sizes (ES)\footnotemark[1] of $0.8$ and $10$.
The magnitudes of all schemes are normalized such that the integral along the angle of departure (AoD) equals unity.
As shown in Fig.~\ref{Fig:electrical size}, when the signal wavelength is comparable to the size of the IRS element (i.e., sub-wavelength regime~\cite{ozdogan2019intelligent} with ES$<$1), the energy of the reflected signal is virtually insensitive to the AoD.
This isotropic distribution leads to a diffuse reflector approximation, which is widely adopted for RF IRS.
However, the magnitude curve becomes sensitive to AoD as the frequency increases (ES$=10$) and transforms into a narrow mainlobe without sidelobes in optical range (ES$>10^4$).
The result reveals that the OIRS reflecting element can be approximated as a specular reflector, where the direction of the mainlobe follows the law of reflection. 
As such, when an OIRS reflecting element is configured to be aligned with a LED/PD pair, it barely interferes with other transceiver antennas, which is referred to as the \textit{angle-selective property} of OIRS.

\subsection{Multi-patch Channel}
\label{Sec:Model multiple}
Based on the single-patch channel, we next develop the general multi-patch model for OIRSs comprising a massive number of reflecting elements.
Let $d_1 = \|\textbf{LR}\|_2$ and $d_2 = \|\textbf{UR}\|_2$ represent the distances between $\textbf{L}$ and $\textbf{R}$ and between $\textbf{R}$ and $\textbf{U}$, respectively.
The RF IRS-reflected channel gain HRF follows a ``multiplicative'' model, which is inversely proportional to the square of the product of $d_1$ and $d_2$ as~\cite{wu2019towards}
\begin{equation}
	\label{Eq:RFirs}
	\boldsymbol{H}^{RF} = \boldsymbol{H}_2\boldsymbol{\Psi}\boldsymbol{H}_1,
\end{equation}
where $\boldsymbol{\Psi} = \text{diag}(\psi_1, \cdots, \psi_N)$ denotes the IRS coefficient matrix with $|\psi_n| \leq 1$, $\boldsymbol{H}_1 \in \mathcal{R}^{N\times N_t}$ denotes the channel matrix between the transmitter and IRS, and $\boldsymbol{H}_2 \in \mathcal{R}^{N_r\times N}$ denotes the channel matrix between the RF IRS and the receiver. 
To overcome the high path loss, the IRS reflection coefficients (amplitude attenuations and phase shifts) of all reflecting elements can be jointly tuned to form an energy concentrated beam and compensate for the multiplicative path loss, referred to as passive beamforming~\cite{gongshimin2020,basar2019wireless,wu2021intelligent,wu2019towards,9771079}.

However, the energy distribution of the lightwave reflected by an OIRS reflecting element is anisotropic due to a large ES. 
Given such angle-selective property, an OIRS with $N$ elements can form $N$ independent reflected beams, based on their alignments with the transceiver antennas.
Define the OIRS coefficient matrices as $\boldsymbol{G} \triangleq [ \boldsymbol{g}_1, \cdots, \boldsymbol{g}_{N_t} ] \in\{0,1\}_{N\times N_t}$ and $\boldsymbol{F} \triangleq [ \boldsymbol{f}_1, \cdots, \boldsymbol{f}_{N_r} ] \in\{0,1\}_{N\times N_r}$.
In particular, $\boldsymbol{g}_{n_t} = [g_{1, n_t}, \cdots, g_{N, n_t}]^T$ and $\boldsymbol{f}_{n_r} = [f_{1, n_r}, \cdots, f_{N, n_r}]^T$, with binary parameters $g_{n, n_t}$ and $f_{n, n_r}$ denoting whether the $n$th OIRS reflecting element is aligned with the $n_t$th LED and the $n_r$th PD, respectively.
Note that each OIRS reflecting element can be aligned with at most one transmitter/receiver antenna, thereby it must hold that $\sum_{n_t=1}^{N_t}g_{n, n_t} \leq 1$ and $\sum_{n_r=1}^{N_r}f_{n, n_r} \leq 1$.
Therefore, the OIRS-reflected channel can be expressed as
\begin{align}
	\label{Eq:channel}
	\boldsymbol{H} &= \left[
	\begin{matrix}
		\left(\boldsymbol{f}_1 \odot \boldsymbol{g}_1\right)^T\boldsymbol{h}_{1,1}  &  \cdots  &  \left(\boldsymbol{f}_1 \odot \boldsymbol{g}_{N_t}\right)^T\boldsymbol{h}_{1,N_t} \\
		\vdots  &  \ddots  &  \vdots \\
		\left(\boldsymbol{f}_{N_r} \odot \boldsymbol{g}_1\right)^T\boldsymbol{h}_{N_r,1}  &  \cdots  &  \left(\boldsymbol{f}_{N_r} \odot \boldsymbol{g}_{N_t}\right)^T\boldsymbol{h}_{N_r,N_t} \\
	\end{matrix}
	\right],
\end{align}
where $\boldsymbol{h}_{n_r,n_t} \in \mathcal{R}_+^{N\times 1}$ denotes the channel gain vector between the $n_t$th LED and the $n_r$th PD.
We assemble $NN_tN_r$ entries of LED-OIRS-PD cascaded channel vectors as $\boldsymbol{H}_c \triangleq [\boldsymbol{h}_{1,1}, \cdots, \boldsymbol{h}_{N_r,1}, \boldsymbol{h}_{1,2}, \cdots, \boldsymbol{h}_{N_r,N_t}]$, which is the CSI matrix that should be estimated.

\subsection{OIRS Channel Estimation Protocol}
\label{Subsec:Model signal}
\begin{figure}[t]
	\centering
	\includegraphics[width=0.48\textwidth]{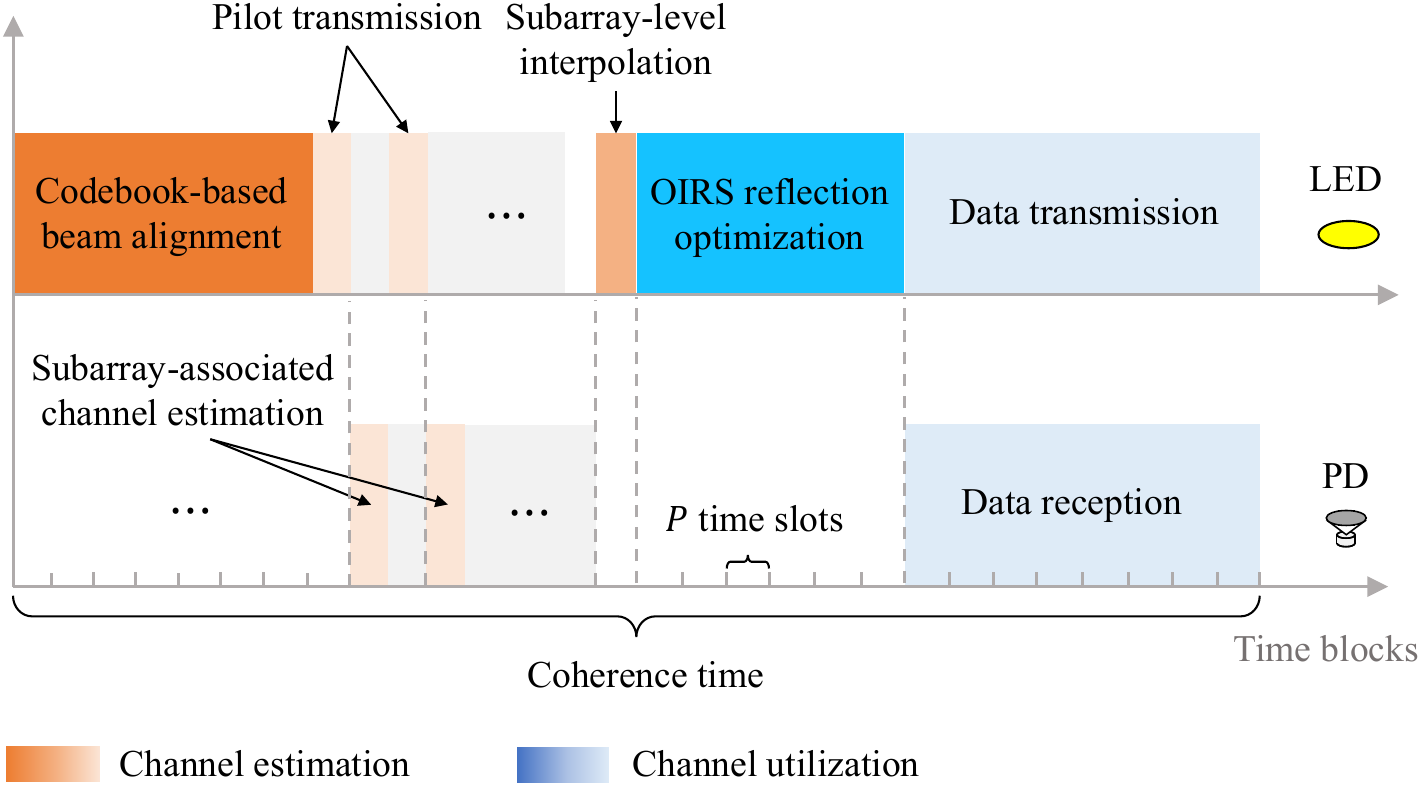}
	\caption{Diagram of the proposed OIRS channel estimation protocol.}
	\label{Fig:protocol}
	\vspace{-0.4cm}
\end{figure}

In this paper, we consider the downlink of an OIRS-assisted MIMO VLC system, which adopts IM/DD scheme under the point source assumption. 
The transmitter is equipped with $N_t$ LEDs and the receiver is equipped with $N_r$ PDs, and an OIRS with $N = N_v \times N_h$ reflecting elements is employed to enhance the performance of VLC. 
We divide the OIRS channel coherence time (to be derived later in Section~\ref{Sec:Coherence}) into multiple blocks, over which the subarrays of the OIRS are sequentially turned on. 
Each time block is further divided into $P$ time slots for pilot transmission to estimate the CSI associated with the ``ON'' subarray. 
In the pth slot of any time block, the received signal is given by
\begin{equation}
	\label{Eq:signal model}
	\boldsymbol{y}_p = \boldsymbol{H}\boldsymbol{x}_p + \boldsymbol{z}_p,
\end{equation}
where $\boldsymbol{x}_p \in \mathcal{R}^{N_t\times 1}_+$, $\boldsymbol{y}_p \in \mathcal{R}^{N_r\times 1}_+$, and $\boldsymbol{z}_p \in \mathcal{R}^{N_r\times 1}$ represent the transmitted pilot signal, the received signal, and the additive white Gaussian noise (AWGN) in this slot, respectively.
By assembling the received signals over all $P$ slots in a time block, the overall received signal can be expressed as
\begin{equation}
	\label{Eq:signal model_cohere}
	\boldsymbol{Y} = \boldsymbol{H} \boldsymbol{X} + \boldsymbol{Z},
\end{equation}
where $\boldsymbol{Y} \triangleq [\boldsymbol{y}_1, \cdots, \boldsymbol{y}_P]$, $\boldsymbol{X} \triangleq [\boldsymbol{x}_1, \cdots, \boldsymbol{x}_P]$, and $\boldsymbol{Z} \triangleq [\boldsymbol{z}_1, \cdots, \boldsymbol{z}_P]$ are stacked matrices.
As such, the main aim of the OIRS channel estimation is to estimate $\boldsymbol{H}_c$ given the received signal $\boldsymbol{Y}$ and the alignment matrices $\boldsymbol{F}$ and $\boldsymbol{G}$ over multiple time blocks.

Despite the extensive existing works for RF IRS channel estimation~\cite{zheng2022survey}, the associated channel estimation methods cannot be applied for OIRS due to the different channel model as given in~(\ref{Eq:channel}). 
In addition, the OIRS channel estimation problem also suffers from the large number of CSI parameters as well as the efficiency of beam alignment. 
In particular, the number of channel parameters to be estimated for OIRS, as per~(\ref{Eq:channel}), is $N N_t N_r$, which far exceeds that in RF IRS~\cite{zheng2022survey}, i.e., $N N_t + N N_r$. 
The increasing parameter number entails more stringent requirement for the complexity of channel estimation. 
Moreover, given the OIRS alignment matrices $\boldsymbol{F}$ and $\boldsymbol{G}$, the rotation angles should be dynamically tuned due to the PDs' movements, which requires accurate and efficient beam sweeping based on the customized codebook.

To address these issues, this paper proposes a new OIRS channel estimation protocol as shown in Fig.~\ref{Fig:protocol}, with the main steps including:
\begin{itemize}
	\item
	\textbf{Codebook-based beam alignment:} 
	Following a geometric optics-based OIRS codebook in the domain of rotational angle, the OIRS reflecting elements are configured to align with specific pairs of transceiver antennas within each channel coherence time. 
	The details will be presented in Section~\ref{Sec:Codebook}.
	
	\item
	\textbf{OIRS-reflected channel estimation:} 
	Based on the received pilot signals given the above beam alignment, the OIRS-reflected channels associated with the aligned transceiver antennas are estimated, followed by an interpolation to retrieve those associated with other transceiver antennas. 
	The OIRS is divided into multiple subarrays for sequential channel estimation to reduce the channel estimation overhead in the above process. 
	The details will be presented in Section~\ref{Sec:Proposed}.
	
	\item 
	\textbf{OIRS reflection optimization and data transmission:} 
	Based on the estimate of the OIRS-reflected channel, the OIRS reflection is optimized at the LED and the optimized reflection design is sent to the OIRS, followed by the data transmission.
\end{itemize}

\section{Theoretical Analysis on OIRS Coherence Characteristics}
\label{Sec:Coherence}
Duo to the unique IM/DD scheme adopted in VLC, which modulates information only on the amplitude of lightwave, the OIRS-reflected channel gain flactuates slowly over different reflecting elements and user locations. 
Such a coherence property is more significant as compared to its counterpart in RF communications and can be exploited for efficient channel estimation. 
In this section, we unveil the spatial coherence and time coherence of the OIRS, as detailed next.

\vspace{-0.1cm}
\subsection{Spatial and Temporal Coherence}
Typically, the Lambertian channel model is widely adopted in IM/DD-based VLC~\cite{komine2004fundamental}, based on which the irradiant intensity reflected by either an optical meta-material-based OIRS or a mirror array-based OIRS is separately derived in~\cite{abdelhady2020visible}. 
For convenience, we present the coherence properties under the single element setting, which corresponds to an arbitrary pair of LED-OIRS-PD. 
Let $\Phi_0$ denote the semi-angle of field-of-view (FOV), and $h(\textbf{R}, \textbf{U})$ denote the OIRS reflected channel gain, where \textbf{R} and \textbf{U} denote the points at the OIRS reflecting element and the detection plane, respectively. 
Then, $h(\textbf{R}, \textbf{U})$ for both of the above two OIRS types are approximately proportional to~\cite{abdelhady2020visible}
\begin{align}
	\label{Eq:lambertian}
	\setlength\abovedisplayskip{3pt}
	h&(\textbf{R}, \textbf{U}) \propto \notag\\
	&\left\{
	\begin{aligned}
		&\frac{\left(\widehat{\boldsymbol{N}}_1^T\widehat{\textbf{LR}}\right)^m\widehat{\boldsymbol{N}}_2^T\widehat{\textbf{UR}}}{\left(\|\textbf{LR}\|_2 + \|\textbf{UR}\|_2\right)^2}, & &\text{if}\ 0 \leq \widehat{\boldsymbol{N}}_2^T\widehat{\textbf{UR}} \leq \cos\Phi_0,   \\
		&0, & &\text{otherwise},
	\end{aligned}
	\right.
	\setlength\belowdisplayskip{3pt}
\end{align}
under the point source assumption. 
Note that $h(\textbf{R}, \textbf{U})$ flactuates with respect to (w.r.t.) two variables: 
(i) Spatial variation caused by the space shift among different OIRS reflecting elements, which is denoted by $\Delta\textbf{R}$; 
(ii) Temporal variation caused by the PD's movement, which is denoted by $\Delta \textbf{U} = \boldsymbol{v}\Delta t$ with $\boldsymbol{v}$ and $\Delta t$ representing the velocity and the the time interval, respectively.

In contrast to the RF channel, the channel gain in VLC barely shows small-scale fading effects due to the IM/DD scheme, i.e., the variation of $h(\textbf{R}, \textbf{U})$ is limited when $\Delta\textbf{R}$ and $\Delta t$ are small, which is termed as the OIRS coherence property. 
Moreover, note that $\Delta\textbf{R}$ is perpendicular to the normal vector of the OIRS surface.

\begin{proposition}
	\label{Prop:Coherence}
	Given a constant $\xi_c \in (0,1)$, the OIRS-reflected channel is $\xi_c$-coherent when $(1-\xi_c)h(\textbf{R}, \textbf{U}) \leq h(\textbf{R} + \Delta\textbf{R}, \textbf{U} + \boldsymbol{v}\Delta t) \leq (1+\xi_c)h(\textbf{R}, \textbf{U})$
\end{proposition}

According to Proposition~\ref{Prop:Coherence}, the channel gain of the OIRS-reflected path can be represented by a single value with an error limitation $\xi_c$. 
However, it is intractable to directly obtain the coherent region since $\Delta\textbf{R}$ and $\Delta t$ are coupled in the channel gain. 
Hence, we adopt an alternate approach and analyze these two variables separately. 
Specifically, let $\xi_s$ denote the relative growth rate of $h(\textbf{R}, \textbf{U})$ w.r.t. $\Delta\textbf{R}$ when $\Delta t = 0$, and $\xi_t$ denote the counterpart w.r.t. $\Delta t$ when $\Delta\textbf{R} = \textbf{0}$. 
Mathematically, $\xi_s$ can be expressed as
\begin{equation}
	\label{Eq:xi_s_definition}
	\xi_s(\Delta\textbf{R}) = \frac{h(\textbf{R} + \Delta\textbf{R}, \textbf{U}) - h(\textbf{R}, \textbf{U})}{h(\textbf{R}, \textbf{U})}.
\end{equation}

\begin{lemma}
	\label{Lemma:growth_rate_space}
	The growth rate of $\xi_s(\Delta\textbf{R})$ is given by
	\begin{align}
		\label{Eq:xi_s}
		\xi_s = &\Big\{ \frac{m}{d_1\cos\theta}\widehat{\boldsymbol{N}}_1 + \frac{1}{d_2\cos\phi}\widehat{\boldsymbol{N}}_2 - (\frac{m}{d_1} + \frac{2}{d_1 + d_2})\widehat{\textbf{LR}} - \notag\\
		&(\frac{1}{d_2} + \frac{2}{d_1 + d_2})\widehat{\textbf{UR}} \Big\}^T \Delta\textbf{R} 
		+ \Delta\textbf{R}^T \Big\{ \frac{-m}{d_1^2\cos^2\theta}\widehat{\boldsymbol{N}}_1\widehat{\boldsymbol{N}}_1^T - \notag\\
		& \frac{1}{d_2^2\cos^2\phi}\widehat{\boldsymbol{N}}_2\widehat{\boldsymbol{N}}_2^T + 2(\frac{m}{d_1^2} + \frac{1}{(d_1 + d_2)^2} + \frac{2}{d_1^2(d_1 + d_2)}) \notag\\
		& \times \widehat{\textbf{LR}}\widehat{\textbf{LR}}^T + 2(\frac{1}{d_2^2} + \frac{1}{(d_1 + d_2)^2}+\frac{2}{d_2^2(d_1 + d_2)})\widehat{\textbf{UR}}\widehat{\textbf{UR}}^T \notag\\
		& + \frac{2}{(d_1 + d_2)^2}(\widehat{\textbf{LR}}\widehat{\textbf{UR}}^T + \widehat{\textbf{UR}}\widehat{\textbf{LR}}^T) - (\frac{m}{d_1^2} + \frac{2}{d_1^2(d_1 + d_2)} \notag\\
		& + \frac{1}{d_2^2} + \frac{2}{d_2^2(d_1 + d_2)})\boldsymbol{I}_3 \Big\} \frac{\Delta\textbf{R}}{2},
	\end{align}
	where $\boldsymbol{I}_N$ denotes an $N \times N$ identity matrix, $\phi$ denotes the
	angle of incidence at the receiver with $\cos(\phi) = \widehat{\boldsymbol{N}}_2^T\widehat{\textbf{UR}}$.
\end{lemma}

The proof of Lemma~\ref{Lemma:growth_rate_space} is provided in Appendix~\ref{app:A}.

Similarly, the relative growth rate of $h(\textbf{R}, \textbf{U})$ w.r.t. $\Delta t$ is expressed as
\begin{equation}
	\label{Eq:xi_t_definition}
	\xi_t(\Delta t) = \frac{h(\textbf{R}, \textbf{U} + \Delta\textbf{U}) - h(\textbf{R}, \textbf{U})}{h(\textbf{R}, \textbf{U})}.
\end{equation}

\begin{lemma}
	\label{Lemma:growth_rate_time}
	The growth rate of $\xi_t(\Delta t)$ is given by
	\begin{align}
		\label{Eq:xi_t}
		\xi_t = &\Big\{ \frac{1}{d_2\cos\phi}\widehat{\boldsymbol{N}}_2 - \left(\frac{1}{d_2} + \frac{2}{d_1 + d_2}\right)\widehat{\textbf{UR}} \Big\}^T \boldsymbol{v}\Delta t \notag\\
		& + \boldsymbol{v}^T \Big\{ 2\left(\frac{1}{d_2^2} + \frac{1}{(d_1 + d_2)^2}+\frac{2}{d_2^2(d_1 + d_2)}\right)\widehat{\textbf{UR}}\widehat{\textbf{UR}}^T -\notag\\
		& \frac{1}{d_2^2\cos^2\phi}\widehat{\boldsymbol{N}}_2\widehat{\boldsymbol{N}}_2^T -\left(\frac{1}{d_2^2} + \frac{2}{d_2^2(d_1 + d_2)}\right)\boldsymbol{I}_3 \Big\} \frac{\boldsymbol{v}(\Delta t)^2}{2}.
	\end{align}
\end{lemma}

The proof of Lemma~\ref{Lemma:growth_rate_time} is provided in Appendix~\ref{app:B}.

\subsection{OIRS Coherence Distance and Time}
\label{Subsec:distance_time}
Based on the aforementioned results, we show that the OIRS coherence condition in Proposition~\ref{Prop:Coherence} is equivalent to the following proposition.
\begin{proposition}
	\label{Prop:Coherence_transform}
	Given a constant $\xi_c \in (0,1)$, the OIRS-reflected channel is $\xi_c$-coherent when $\max(|\xi_s(\Delta\textbf{R})|, |\xi_t(\Delta t)|) \leq \xi_c$.
\end{proposition}

It follows from Proposition~\ref{Prop:Coherence_transform} that the coherent region of the OIRS-reflected channel gain can be obtained separately. 
Specifically, in the temporal domian, the feasible space can be expressed as $|\xi_t(\Delta t)| \leq \max(|\xi_s(\Delta\textbf{R})|, |\xi_t(\Delta t)|) \leq \xi_c$; thereby, the OIRS coherence time tc is defined as the largest time interval as
\begin{align}
	\label{Eq:t_c_problem}
	t_c \triangleq \ & \Delta t_{\max} - \Delta t_{\min} \\
	\label{Eq:t_c_problem_constraint}
	&\text{s.t.}\ |\xi_t(\Delta t)| \leq \xi_c.
\end{align}
However, in the spatial domain, the feasible space of $\Delta\textbf{R}$ is given by $|\xi_s(\Delta\textbf{R})| \leq \max(|\xi_s(\Delta\textbf{R})|, |\xi_t(\Delta t)|) \leq \xi_c$, which results in a two-dimentional region. 
Thus, the OIRS coherence distance dc refers to the minimal value of the maximum length within the feasible region, which can be expressed as
\begin{align}
	\label{Eq:d_c_problem}
	d_c \triangleq \min\limits_{\begin{subarray}{c}
			\Delta\textbf{R}_1, \Delta\textbf{R}_2
	\end{subarray}} & \max\ \|\Delta\textbf{R}_1\|_2 + \|\Delta\textbf{R}_2\|_2 \\
	\text{s.t.}\ &|\xi_s(\Delta\textbf{R}_1)| \leq \xi_c,\\
	& |\xi_s(\Delta\textbf{R}_2)| \leq \xi_c,\\
	& \Delta\textbf{R}_2 = -c \Delta\textbf{R}_1, \exists c>0.
\end{align}

\begin{thm}
	\label{Theorem:coherence_distance}
	Given $\xi_c \in (0,1)$, the OIRS coherence time $t_c$ is given by
	\begin{align}
		\label{Eq:t_c}
		t_c = \left\{
		\begin{aligned}
			&|\Delta t_2|\sqrt{1+\frac{4\xi_c /|\Delta t_2|}{\left|\Big\{\frac{1}{d_2\cos\phi}\widehat{\boldsymbol{N}}_2 - \left(\frac{1}{d_2} + \frac{2}{d_1 + d_2}\right)\widehat{\textbf{UR}} \Big\}^T \boldsymbol{v} \right|}}, \\
			& \qquad\qquad\qquad\qquad\qquad\qquad\qquad \text{if}\ \left|\xi_t\left(\frac{\Delta t}{2}\right)\right| \leq \xi_c,   \\
			&\frac{2\xi_c}{\left|\Big\{\frac{1}{d_2\cos\phi}\widehat{\boldsymbol{N}}_2 - \left(\frac{1}{d_2} + \frac{2}{d_1 + d_2}\right)\widehat{\textbf{UR}} \Big\}^T \boldsymbol{v} \right|},\quad \text{otherwise},
		\end{aligned}
		\right.
	\end{align}
	where $\Delta t_2$ is obtained by~(\ref{Eq:delta_t2}).
\end{thm}

\begin{figure*}[b]
	{\noindent} \rule[-5pt]{18cm}{0.05em}
	\begin{equation}
		\label{Eq:delta_t2}
		\Delta t_2 = \frac{2\Big\{\frac{1}{d_2\cos\phi}\widehat{\boldsymbol{N}}_2 - \left(\frac{1}{d_2} + \frac{2}{d_1 + d_2}\right)\widehat{\textbf{UR}} \Big\}^T \boldsymbol{v}}{\boldsymbol{v}^T \Big\{ \left(\frac{1}{d_2^2} + \frac{2}{d_2^2(d_1 + d_2)}\right)\boldsymbol{I}_3 + \frac{1}{d_2^2\cos^2\phi}\widehat{\boldsymbol{N}}_2\widehat{\boldsymbol{N}}_2^T - 2\left(\frac{1}{d_2^2} + \frac{1}{(d_1 + d_2)^2}+\frac{2}{d_2^2(d_1 + d_2)}\right)\widehat{\textbf{UR}}\widehat{\textbf{UR}}^T \Big\} \frac{\boldsymbol{v}(\Delta t)^2}{2}}
	\end{equation}
\end{figure*}

The proof of Theorem~\ref{Theorem:coherence_distance} is provided in Appendix~\ref{app:C}.

Let $\boldsymbol{w}$ denote the coefficient vector of $\Delta\textbf{R}$, and $\boldsymbol{W}$ represent the coefficient matrix in~(\ref{Eq:xi_s}) such that $\xi_s = \boldsymbol{w}^T\Delta\textbf{R} + \Delta\textbf{R}^T\boldsymbol{W}\Delta\textbf{R}$. 
Then, we have the following theorem.

\begin{thm}
	\label{Theorem:coherence_time}
	Given $\xi_c \in (0,1)$, the OIRS coherence distance $d_c$ is given by
	\begin{equation}
		\label{Eq:d_c_expression}
		d_c = \min\limits_{\begin{subarray}{c}
				\widehat{\Delta\textbf{R}}
		\end{subarray}} d_c(\widehat{\Delta\textbf{R}}),
	\end{equation}
	where $d_c(\widehat{\Delta\textbf{R}})$ can be obtained as
	\begin{align}
		\label{Eq:d_c}
		d_c = \left\{
		\begin{aligned}
			&|\Delta r_2|\sqrt{1+\frac{4\xi_c /|\Delta r_2|}{\left| \boldsymbol{w}^T\widehat{\Delta\textbf{R}} \right|}},\quad \text{if}\ \left|\xi_s\left(\frac{\Delta r_2\widehat{\Delta\textbf{R}}}{2}\right)\right| \leq \xi_c,   \\
			&\frac{2\xi_c}{\left| \boldsymbol{w}^T\widehat{\Delta\textbf{R}} \right|},\qquad\qquad\qquad\ \text{otherwise},
		\end{aligned}
		\right.
	\end{align}
	where $\Delta r_2$ is given by
	\begin{equation}
		\label{Eq:delta_r2}
		\Delta r_2 = \frac{-\boldsymbol{w}}{\widehat{\Delta\textbf{R}}^T\boldsymbol{W}\widehat{\Delta\textbf{R}}}.
	\end{equation}
\end{thm}

The proof of Theorem~\ref{Theorem:coherence_time} is provided in Appendix~\ref{app:D}.

It follows from Theorem~\ref{Theorem:coherence_time} that the coherence distance $d_c$ can be obtained by calculating~(\ref{Eq:d_c}) with $\widehat{\Delta\textbf{R}} = [1, 0, 0]$ and $\widehat{\Delta\textbf{R}} = [0, 0, 1]$, which constitute an orthogonal basis of the space shift $\widehat{\Delta\textbf{R}}$ on OIRS. 
By leveraging the OIRS coherence distance $d_c$ in~(\ref{Eq:d_c}) and OIRS coherence time $t_c$ in~(\ref{Eq:t_c}), the OIRS channel estimation can be significantly improved, as shown next.

\begin{figure}[t]
	\centering  
	\subfigure[Uniform OIRS codebook]{   
		\centering    
		\includegraphics[width=0.22\textwidth]{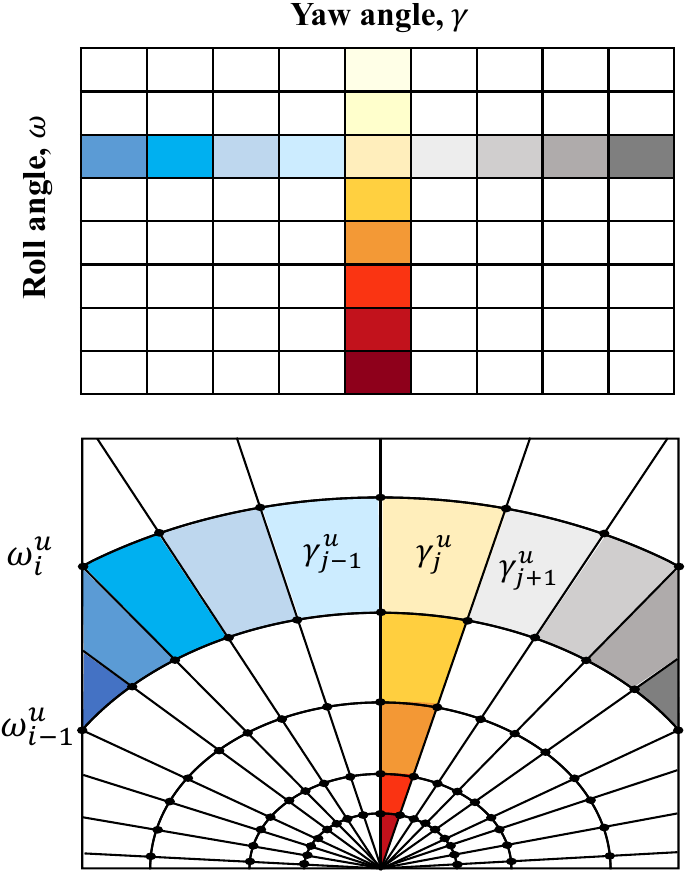}
		\label{Fig: uniform codebook}
	}
	\hspace{-0.1cm}
	\subfigure[Non-uniform OIRS codebook]{ 
		\centering    
		\includegraphics[width=0.23\textwidth]{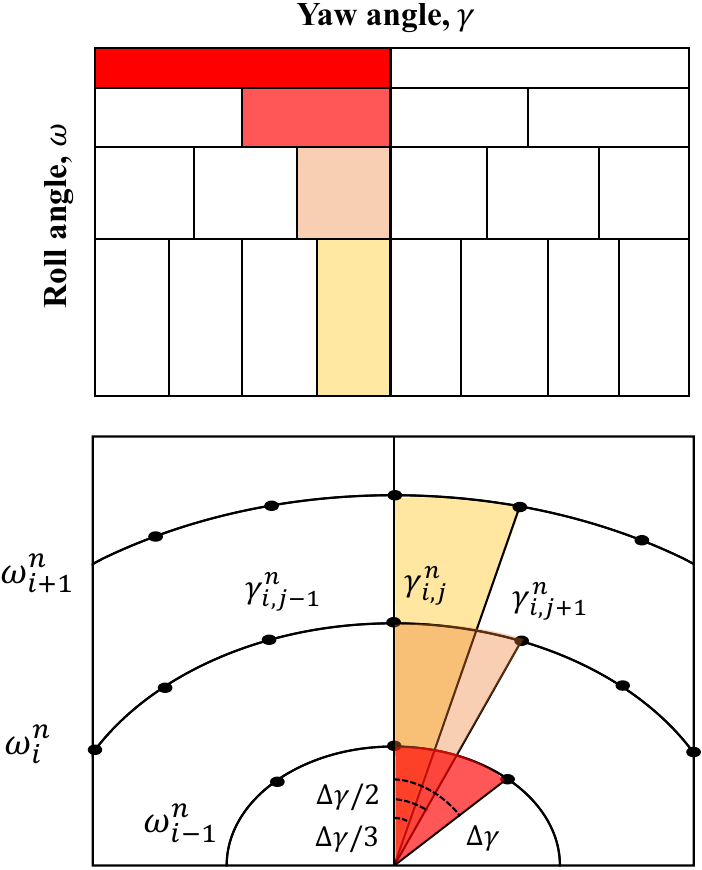}
		\label{Fig: nonuniform codebook}
	}
	\caption{Comparison between different OIRS codebook designs and distributions of codewords in the detection plane.}    
	\label{Fig: codebook design}    
\end{figure}


\vspace{-0.1cm}
\section{OIRS Codebook Design}
\label{Sec:Codebook}
In this section, we propose a dedicated rotation-angle codebook design based on the mirror array OIRS to achieve fast beam alignment within each channel coherence time, whereas that for the meta-surface-based OIRS can be similarly achieved.

Specifically, for a pair of OIRS reflecting element and LED, the OIRS-reflected channel gain at location \textbf{P} can be expressed as
\begin{align}
	\label{Eq:channel gain}
	\setlength\abovedisplayskip{3pt}
	h_{\textbf{P}}(\omega, \gamma) = \iint_{\textbf{P} \in \mathcal{U}} E(\textbf{P}; \omega, \gamma) d\textbf{P}_x d\textbf{P}_y,
	\setlength\belowdisplayskip{3pt}
\end{align}
w.r.t. the roll angle $\omega$ and the yaw angle $\gamma$ of the element.
Let $\mathcal{W} \triangleq \left\{\omega_1, \cdots, \omega_{|\mathcal{W}|}\right\}$ and $\mathcal{Y} \triangleq \left\{\gamma_1, \cdots, \gamma_{|\mathcal{Y}|}\right\}$ denote the roll angle codebook and the yaw angle codebook, respectively.
Hence, the cartesian product of $\mathcal{W}$ and $\mathcal{Y}$, namely $\{(\omega, \gamma)|\omega\in\mathcal{W}, \gamma\in\mathcal{Y}\}$, gives rise to the OIRS codebook.
Generally, the codebook design aims to maximize the expected channel gain~\cite{wang2023hierarchical,zhang2017codebook}, namely $\max_{\mathcal{W}, \mathcal{Y}} \mathbb{E}_{\textbf{U}} [h_{\textbf{U}}(\omega, \gamma)]$, by jointly optimizing the codebooks $\mathcal{W}$ and $\mathcal{Y}$. 
Therefore, the optimization problem of the OIRS codebook design can be expressed as
\vspace{-0.2cm}
\begin{align}
	&\mathop{\arg\max} \limits_{\begin{subarray}{c}
			\mathcal{W}, \mathcal{Y}
	\end{subarray}}  \ \mathbb{E}_{\textbf{U}} [h_{\textbf{U}}(\omega, \gamma)] \label{objective}	\\
	&\ \text{s.t.}\ \omega \in \mathcal{W}, \\
	&\qquad \gamma \in \mathcal{Y}, \\
	&\qquad \max\limits_{\begin{subarray}{c}
			\mathcal{W}, \mathcal{Y}
	\end{subarray}} \ h_{\textbf{P}}(\omega, \gamma) > 0,\ \forall \textbf{P},	\label{Eq:codebook_constriant}
\end{align}
where the constraint in~(\ref{Eq:codebook_constriant}) guarantees that the whole feasible space of \textbf{P} is covered by the angle codewords. 
However, this problem is difficult to solve due to the complicated expression in~(\ref{Eq:channel gain}).

A straightforward solution is to generate $\mathcal{W}$ and $\mathcal{Y}$ by uniformly sampling the interval $[-\pi/2, \pi/2)$.
Let $\Delta \omega^u$ and $\Delta \gamma^u$ denote the sampling spacing in the roll-angle and yaw-angle domains, respectively.
As such, the roll angle codebook can be obtained as
\begin{equation}
	\label{Eq:uniform codebook roll}
	\omega^u_i = \pm i \Delta \omega^u,\ \ i=1, 2, \cdots, |\mathcal{W}|,
\end{equation}
and the yaw angle codebook is given by
\begin{equation}
	\label{Eq:uniform codebook yaw}
	\gamma^u_j = \pm j \Delta \gamma^u,\ \ j=1, 2, \cdots, |\mathcal{Y}|.
\end{equation}
Such a uniform OIRS codebook is intuitive and simple to implement; however, it incurs practically exorbitant overhead for OIRS channel estimation. 
As depicted in Fig.~\ref{Fig: uniform codebook}, when $\omega$ is fixed and $\gamma$ is varied, the angle codewords cover adjacent regions within a single ring. 
When $\omega$ is increased, the angle codewords cover the regions within several concentric rings with an increasing radius. 
It can be observed that the angle codewords in the inner ring are highly overlapped, indicating that the uniform sampling strategy may lead to a severely imbalanced distribution of the angle codewords.

To tackle this issue, we propose a geometric optics (GO)-based OIRS codebook as depicted in Algorithm~\ref{Algo:codebook}, to realize a uniform distribution of its constituent angle codewords in the detection plane by non-uniformly sampling the roll angle $\omega$ and yaw angle $\gamma$.

\textit{Roll angle sampling:} First, considering the above condition, a constant incremental radius is requires when the roll angle changes. 
Let $\gamma$ be initialized as
\begin{equation}
	\label{Eq:gamma_c}
	\gamma_c = \arctan \left( \frac{\boldsymbol{e}_2^T\widehat{\textbf{LR}}}{\boldsymbol{e}_1^T\widehat{\textbf{LR}}} \right),
\end{equation}
such that $\widehat{\boldsymbol{N}}(\omega, \gamma_c)$ and $\widehat{\boldsymbol{N}}_1$ are coplanar.
Let $\omega_{i}^n$ denote the $i$th roll-angle codeword. 
Based on the principles of geometric optics, the $(i + 1)$th roll-angle codeword can be determined as $\!\!\!\!\!\!$
\begin{equation}
	\label{Eq:omega sampling}
	\setlength\abovedisplayskip{3pt}
	\tan(\alpha + 2 \omega_{i+1}^n) + \tan(\alpha + 2 \omega_{i-1}^n) = 2\tan(\alpha + 2 \omega_{i}^n),
	\setlength\belowdisplayskip{3pt}
\end{equation}
where $\alpha \triangleq \arccos(\boldsymbol{e}^T_3\widehat{\textbf{LR}})$.
The first codeword $\omega_{1}^n$ can be initialized to direct the incident signal vertically downwards, which is given by
\begin{equation}
	\label{Eq:omega_1}
	\omega_{1}^n = - \arccos \left( \frac{\left( \widehat{\textbf{LR}} + \boldsymbol{e}_3 \right)^T \widehat{\boldsymbol{N}}(0, \gamma_c)}{\| \widehat{\textbf{LR}} + \boldsymbol{e}_3 \|_2} \right).
\end{equation}
Furthermore, $\omega_{2}^n$ is given by $\omega_{2}^n = \omega_{1}^n + \Delta \omega^n$ with $\Delta \omega^n$ being a fixed constant.
As such, the proposed roll angle codebook can be obtained recursively by repeating the process in~(\ref{Eq:omega sampling}).

\textit{Yaw angle sampling:} On the other hand, the uniform codeword condition requires a constant arc length. 
To this end, the yaw angle codebook with $\omega_{i}^n$ should be generated based on the uniform sampling, since the arc length is proportional to $\gamma$. 
However, considering the radius for $\omega = \omega_{i}^n$ equals that for $\omega = \omega_{1}^n$ times $i$, the sample spacing of $\gamma$ should be $\Delta \gamma^n / i$ in the $i$th circle. 
It follows that the yaw angle when $\omega = \omega_{i}^n$ should be
\begin{equation}
	\label{Eq:gamma sampling}
	\gamma_{i, j}^n = \gamma_c \pm j\frac{\Delta \gamma^n}{i},\ \ j=1, 2, \cdots. 
\end{equation}

As shown in Fig.~\ref{Fig: nonuniform codebook}, the codewords of the proposed non-uniform OIRS codebook are uniformly distributed in the detection plane, so that the number of codewords can be significantly reduced. 
Therefore, the fast beam alignment can be realized for improving the efficiency of the channel estimation of the OIRS-assisted VLC system.

\begin{algorithm}[t]
	\caption{GO-based Non-uniform OIRS Codebook}
	\label{Algo:codebook}
	\textbf{Input:} $\Delta \omega^n$, $\Delta \gamma^n$\\
	\textbf{Output:} $\mathcal{W}$, $\mathop{\cup}\limits_{i} \mathcal{Y}_i$
	\begin{algorithmic}[1]
		\State Calculate $\gamma_c$ based on~(\ref{Eq:gamma_c}) and set $\gamma \gets \gamma_c$;
		\State Calculate $\omega_1$ based on~(\ref{Eq:omega_1}) and set $\omega \gets \omega_1$;
		\State Calculate $\omega_2 \gets \omega_1 + \Delta \omega^n$;	
		\State Set $\mathcal{W} \gets \{\omega_1,\ \omega_2\}$;
		\State Set $\mathcal{Y}_1 \gets \{\gamma_c\}$ and $\mathcal{Y}_i \gets \emptyset,\ \forall i > 1$;
		\State Set $i \gets 2$;
		\Repeat
		\Repeat
		\State Reduce sampling interval to $\Delta \gamma^n / i$;
		\State Calculate $\gamma_{i, j}$ based on~(\ref{Eq:gamma sampling});
		\State Set $\mathcal{Y}_i \gets \mathcal{Y}_i \cup \{\gamma_{i,j}\}$;
		\State Update the index: $j \gets j + 1$;
		\Until{$|\gamma_{i, j}| > \pi/2$}
		\State Calculate $\omega_{i+1}^n$ based on~(\ref{Eq:omega sampling});
		\State Set $\mathcal{W} \gets \mathcal{W} \cup \{\omega_{i+1}^n\}$;
		\State Set $\omega_{i-1}^n \gets \omega_{i}^n$ and $\omega_{i}^n \gets \omega_{i+1}^n$;
		\State Update the index: $i \gets i + 1$;
		\Until{\textit{Out-of-bounds.}}
	\end{algorithmic}
\end{algorithm}
\setlength{\textfloatsep}{0.5cm}

\section{Proposed JSTS Algorithm for\\ OIRS Channel Estimation}
\label{Sec:Proposed}
In this section, we propose a sequential channel estimation method to estimate the OIRS-reflected VLC channel within each coherence time for the aligned transceiver pairs first. 
The full CSI for those non-aligned transceiver pairs is then acquired by performing a spatial interpolation.

\subsection{Sequential Channel Estimation}
Specifically, we replace the OIRS alignment matrices $\boldsymbol{G}$ and $\boldsymbol{F}$ equivalently by a single matrix $\boldsymbol{V} \in\{0,1\}^{N\times N_tN_r}$, wherein each column is given by~\cite{10190313}
\begin{equation}
	\label{Eq:v generate}
	\boldsymbol{v}_{n_r + (n_t - 1)N_r} = \boldsymbol{f}_{n_r} \odot \boldsymbol{g}_{n_t}.
\end{equation}
Due to the constraints that $\sum_{n_t=1}^{N_t}g_{n, n_t} \leq 1$ and $\sum_{n_r=1}^{N_r}f_{n, n_r} \leq 1$, we have $\sum_{q=1}^{N_tN_r}v_{n, q} \leq 1$.
Let $\textit{blkdiag}(\boldsymbol{A})$ denote the block diagonal matrix with its $i$th main diagonal element be the $i$th column vector of $\boldsymbol{A}$. 
The OIRS-reflected channel in~(\ref{Eq:channel}) can be recast as
\begin{equation}
	\label{Eq:MIMO_Vlinear}
	\text{vec}\left( \boldsymbol{H} \right) = \textit{blkdiag}\left( \boldsymbol{V} \right)^T\text{vec}\left(\boldsymbol{H}_c\right).
\end{equation}
According to the equation $\text{vec}(\boldsymbol{A}\boldsymbol{B}\boldsymbol{C}) = (\boldsymbol{C}^T \otimes \boldsymbol{A}) \text{vec}(\boldsymbol{B})$~\cite{horn1994topics}, the received signal in~(\ref{Eq:signal model_cohere}) can be vectorized as
\begin{align}
	\label{Eq:observed signal}
	\text{vec}\left(\boldsymbol{Y}\right) &= \left(\boldsymbol{X}^{T} \otimes \boldsymbol{I}_{N_r}\right)\text{vec}\left(\boldsymbol{H}\right) + \text{vec}\left(\boldsymbol{Z}\right) \notag\\
	&= \left(\boldsymbol{X}^{T} \otimes \boldsymbol{I}_{N_r}\right) \textit{blkdiag}\left(\boldsymbol{V}\right)^T\text{vec}\left(\boldsymbol{H}_c\right) + \text{vec}\left(\boldsymbol{Z}\right).
\end{align}
Therefore, given the pilot matrix $\boldsymbol{X}$, the received signal matrix $\boldsymbol{Y}$, and the OIRS reflection pattern $\boldsymbol{V}$, the cascaded channel matrix $\boldsymbol{H}_c$ can be estimated using conventional estimation methods such as the MMSE estimator based on~(\ref{Eq:observed signal}).

Nonetheless, at most $N_tN_r$ entries of $\boldsymbol{H}_c$ can be estimated each time due to the rank-deficit property of the matrix $(\boldsymbol{X}^{T} \otimes \boldsymbol{I}_{N_r}) \textit{blkdiag}(\boldsymbol{V})^T$, i.e., 
\begin{align}
	\label{Eq:low-rank}
	& \rank\left( (\boldsymbol{X}^{T} \otimes \boldsymbol{I}_{N_r})  \textit{blkdiag}(\boldsymbol{V})^T \right) \notag\\
	&\qquad \overset{(a)}{\leq}  \min \left( \rank(\boldsymbol{X}^{T} \otimes \boldsymbol{I}_{N_r}), \rank(\textit{blkdiag}(\boldsymbol{V})) \right) \notag\\
	&\qquad \overset{(b)}{\leq} N_t N_r,
\end{align}
where $(a)$ is due to $\rank(\boldsymbol{A}\boldsymbol{B}) \le \min{\rank(\boldsymbol{A}), \rank(\boldsymbol{B})}$~\cite{horn1994topics} and $(b)$ is due to the fact that $\rank(\boldsymbol{X}^{T} \otimes \boldsymbol{I}_{N_r}) = \rank(\boldsymbol{X}) \times \rank(\boldsymbol{I}_{N_r}) \leq N_t N_r$.
It follows that the complete CSI paramters of the OIRS-reflected channel requires a minimum of $NN_tN_r / \rank((\boldsymbol{X}^{T} \otimes \boldsymbol{I}_{N_r})  \textit{blkdiag}(\boldsymbol{V})^T) \geq N$ estimations, which result in high overhead for practical OIRS channel estimation.

In light of this, unlike the IRS training patterns used in RF~\cite{mishra2019channel,zheng2020intelligent,jensen2020optimal}, we design a dedicated OIRS reflection pattern to exploit the joint space-time coherence of the OIRS-reflected VLC channel. 
As depicted in Fig.~\ref{Fig:sample}, considering the OIRS spatial coherence, the OIRS reflecting elements are divided into $Q = Q_v \times Q_h$ subarrays each with an aperture of $d_c$. 
As such, there are $(\lceil d_c/b \rceil)^2$ elements in each subarray, where the corresponding channel parameters between a specific pair of LED and PD can be represented by one parameter. 
For the OIRS reflecting element situated at the $n_r$th row and the $n_t$th column of the $q$th subbarray, with $q=(q_v-1)Q_h+q_h \in \{1,\cdots,Q\}$, $q_v\in{1, \cdots,Q_v}$, $q_h\in{1, \cdots,Q_h}$, its row index and column index are given by
\begin{align}
	i &= (q_v - 1) \left\lceil \frac{d_c}{b} \right\rceil + n_r, \label{Eq:i} \\
	j &= (q_h - 1) \left\lceil \frac{d_c}{b} \right\rceil + n_t. \label{Eq:j}
\end{align}
As such, the index of this OIRS reflecting element is obtained as $n=(i-1)N_h+j$ and its alignment with the LED/PD pairs can be configured as
\begin{equation}
	\label{Eq:reflection pattern}
	f_{n, n_r}^* = g_{n, n_t}^* = 1.
\end{equation}
We denote the resulting alignment matrix by $\boldsymbol{G}^*$ and $\boldsymbol{F}^*$, and its equivalent OIRS coefficient matrix is denoted by $\boldsymbol{V}^*$ according to~(\ref{Eq:v generate}).

\begin{figure}[t]
	\centering
	\includegraphics[width=0.47\textwidth]{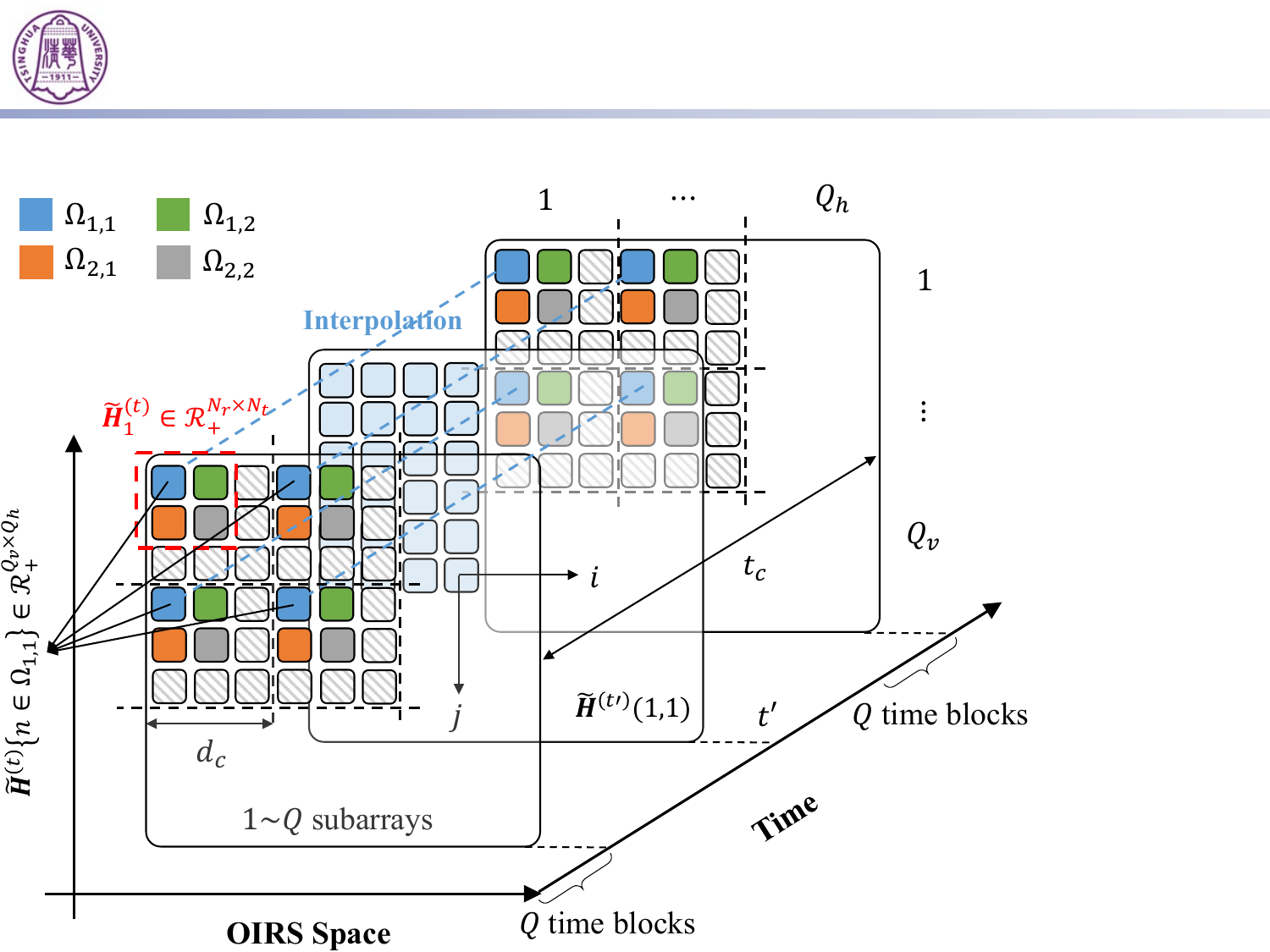}
	\caption{Sequential OIRS channel estimation.}
	\label{Fig:sample}
\end{figure}

Next, the OIRS's $Q$ subarrays are activated sequentially over $Q$ time blocks. 
Without loss of generality, when the $q$th subarray is activated, we consider that the rows of $\boldsymbol{V}^*$ that have the same indices as the OIRS reflecting elements in this subarray remain unchanged, while other rows are set to zeros. 
We denote the resulting matrix $\boldsymbol{V}^*$ as $\boldsymbol{V}^*_q$ and the effective channel at time block $q$ can be expressed as
\begin{equation}
	\text{vec}\left( \boldsymbol{H}_q \right) = \textit{blkdiag}\left(\boldsymbol{V}^*_q \right)^T\text{vec}\left(\boldsymbol{H}_c\right),
\end{equation}
based on~(\ref{Eq:MIMO_Vlinear}).
The corresponding received signal is given by
\begin{equation}
	\label{Eq:local_signal_model}
	\boldsymbol{Y}_q^{(t)} = \boldsymbol{H}_q \boldsymbol{X}_q^{(t)} + \boldsymbol{Z}_q^{(t)}.
\end{equation}
For~(\ref{Eq:local_signal_model}), the MMSE estimator can be utilized for the channel estimation of the $q$th subarray, which is given by
\begin{equation}
	\label{Eq:MMSE estimator}
	\boldsymbol{\tilde{H}}^{(t)}_q = \boldsymbol{Y}^{(t)}_q\boldsymbol{X}^{(t)T}_q(\boldsymbol{X}^{(t)}_q\boldsymbol{X}^{(t)T}_q + \sigma^2 \boldsymbol{I}_{N_t})^{-1}.
\end{equation}
Note that $\rank(\boldsymbol{V}^{(k)}_*) = N_tN_r$ under the designed OIRS reflection pattern. 
Hence, there are $N_tN_r$ entries of $\boldsymbol{H}_c$ estimated in $\boldsymbol{\tilde{H}}^{(t)}_q$, where $[\boldsymbol{\tilde{H}}^{(t)}_q]_{n_r,n_t}$ represents the cascaded channel gain from the $n_t$th LED to the $n_r$th PD aligned by the specific reflecting element in the $q$th subarray. 
Furthermore, let $b$ denote the spacing between two adjacent OIRS reflecting elements, as shown in Fig.~\ref{Fig:multiple}. 
Then the number of CSI parameters to be estimated is reduced by a factor of $d_c^2/b^2$.

\begin{algorithm}[t]
	\caption{Proposed JSTS Algorithm}
	\label{Algo:sampling}
	\textbf{Input:} $N_t$, $N_r$, $N$, $\mathcal{W}$, $\mathop{\cup}\limits_{i} \mathcal{Y}_i$, $t_c$, $d_c$\\
	\textbf{Output:} $\boldsymbol{G}^*$, $\boldsymbol{F}^*$, $\tilde{\boldsymbol{H}}_c$
	\begin{algorithmic}[1]
		\State Set OIRS reflection pattern $\boldsymbol{G}^*$ and $\boldsymbol{F}^*$ based on~(\ref{Eq:i})-(\ref{Eq:reflection pattern});
		\While{$\Delta t > t_c$}
		\For{$n = 1$ to $N$}	\Comment{Fast beam alignment} 
		\State Set $n_t \gets \arg\max\boldsymbol{G}_{n, :}$;
		\State Set $n_r \gets \arg\max\boldsymbol{F}_{n, :}$;
		\State Fast beam alignment: $\omega^* \in \mathcal{W}$, $\gamma^* \in \mathcal{Y}_i$;
		\State Configure the $n$th OIRS reflecting element by $\omega^*$ and $\gamma^*$;
		\EndFor
		\State Determine $Q_v$ and $Q_h$ based on $d_c$;
		\For{$o = 1$ to $Q_vQ_h$}	\Comment{Sequential estimation} 
		\State Activate the $q$th subarray with setting $\boldsymbol{V} = \boldsymbol{V}_q^*$;
		\State Estimate $\boldsymbol{H}^{(t)}_q$ based on~(\ref{Eq:MMSE estimator});
		\EndFor
		\State Obtain CSI parameters $\boldsymbol{\tilde{H}}^{(t)}_{n\in \Omega_{n_r, n_t}}$ based on~(\ref{Eq:group cup});
		\For{$n_r = 1$ to $N_r$}	\Comment{Space-time interpolation}
		\For{$n_t = 1$ to $N_t$}
		\State Perform the interpolation to obtain $\boldsymbol{\tilde{H}}^{(t')}_{n_r, n_t}$
		\Statex \qquad \qquad\ based on~(\ref{Eq:interpolation});
		\State Estimate $\tilde{\boldsymbol{h}}_{n_r,n_t}$ as 
		$\tilde{\boldsymbol{h}}_{n_r,n_t} \gets \text{vec}(\boldsymbol{\tilde{H}}^{(t')}_{n_r, n_t})$;
		\EndFor
		\EndFor
		\State Obtain CSI matrix as $\tilde{\boldsymbol{H}}_c = [\tilde{\boldsymbol{h}}_{1,1}, \cdots, \tilde{\boldsymbol{h}}_{N_r,N_t}]$;
		\EndWhile
	\end{algorithmic}
\end{algorithm}
\setlength{\textfloatsep}{0.5cm}

\vspace{-0.1cm}
\subsection{Subarray-wise Interpolation}
\vspace{-0.1cm}
The full CSI matrix Hc (including the CSI for non-aligned transceiver antennas) can be retrieved by exploiting interpolation across different subarrays and time blocks. 
For brevity, we focus on the CSI reconstruction between the $n_t$th LED and the $n_r$th PD, while it can be similarly applied to other pairs of LED and PD. 
Let $\Omega_{n_r, n_t}$ denote the index set of the OIRS reflecting elements aligned with the $n_t$th LED and the $n_r$th PD, i.e.,
\begin{equation}
	\label{Eq:cluster}
	\Omega_{n_r, n_t} \triangleq \{n| f_{n, n_r} = g_{n, n_t} = 1, n\in \{1,\cdots,N\}\},
\end{equation}
which are marked by a unique color in Fig.~\ref{Fig:sample} for illustration.
By collecting $[\boldsymbol{\tilde{H}}^{(t)}_q]_{n_r,n_t}$ across all subarrays, the estimated CSI parameters at time block $q$ can be obtained as
\begin{equation}
	\label{Eq:group cup}
	\boldsymbol{\tilde{H}}^{(t)}_{n\in \Omega_{n_r, n_t}} = \mathop{\cup}\limits_{q} \left\{[\boldsymbol{\tilde{H}}^{(t)}_q]_{n_r,n_t}\right\},
\end{equation}
which is of the size $Q_v \times Q_h$, and the entry $[\boldsymbol{\tilde{H}}^{(t)}\{n\in \Omega_{n_r, n_t}\}]_{q_v,q_h}$ denotes the estimated channel gain of the OIRS reflecting element at its $i$th row and $j$th column, as expressed in~(\ref{Eq:i}) and~(\ref{Eq:j}), respectively. 
Therefore, by leveraging the spatial and temporal coherences of the OIRS-reflected channel, the channel gain for the $[(i-1)N_h + j]$th reflecting element can be obtained by a three-dimensional interpolation as
\begin{align}
	\label{Eq:interpolation}
	[\boldsymbol{\tilde{H}}^{(t')}_{n_r, n_t}]_{i, j} = \sum_{t}\sum_{q_v}\sum_{q_h} w(t', t; i, q_v; j, q_h) [\boldsymbol{\tilde{H}}^{(t)}_{n\in \Omega_{n_r, n_t}}]_{q_v,q_h},
\end{align}
at the $p$th time slot, where the weight function $w(\cdot)$ can be cubic, spline, etc~\cite{hong2012novel}. 
The proposed JSTS algorithm for OIRS channel estimation is summarized in Algorithm~\ref{Algo:sampling}.

\section{Numerical Results}
\label{Sec:numerical}
In this section, we provide numerical results to evaluate the performance of our proposed non-uniform codebook and OIRS channel estimation algorithm. 
We consider a point-to-point VLC setting in an indoor environment with dimensions of 4m$\times$4m$\times$3m. 
The uniform planar array (UPA) transmitter, which is fixed on the ceiling, is composed of multiple LEDs that are in the shape of circles with the radius of 0.1m. 
For each single LED, its Lambertian index, the FOV, and the optical filter gain are set to $1$, $70\degree$, and $1$, respectively. 
The OIRS is composed of a set of rectangular patches in the size of $a = 0.05\text{m}$ and the spacing between adjacent elements is $b = 0.1\text{m}$. 
Its center coordinate is set to (2m, 0m, 1.5m) and the reflectivity of each reflecting element is $\delta= 0.9$. 
The multi-PD receiver is also equipped with a UPA, where each PD is in the shape of a rectangle with a length of 0.1m, and the spacing between two PDs is 0.4m. 
Assuming the direct LoS path is obstructed, the emitted light can be received via the reflection from the OIRS mounted on the wall. 
The OIRS-reflected channel is generated by the optical simulation according to~(\ref{Eq:channel gain}).

First, we consider a simplified single-input single-output (SISO) VLC system to evaluate the performance of the proposed GO-based non-uniform OIRS codebook. 
Specifically, we set $N = 1$ and $N_t = N_r = 1$, where a single-LED transmitter is located at (2m, 2m, 3m). 
According to Algorithm~\ref{Algo:codebook}, the proposed non-uniform OIRS codebook can be generated by two given parameters $\Delta \omega^n$ and $\Delta \gamma^n$. 
For any given OIRS coefficient matrix $\boldsymbol{V}$, the IMA physical parameters of the element (namely $\omega$ and $\gamma$) can be obtained by beam sweeping within the codebook $\mathcal{W}$ and $\mathop{\cup}\limits_{i} \mathcal{Y}_i$.
Particularly, it is reasonably assumed that a coarse location of the target PD is known by a certain VLC positioning technique~\cite{zhuang2018survey}, and we use a circle with radius of $r$ to represent the positioning precision. 
As such, the beam alignment of the OIRS reflecting elements can be improved by sweeping a subset of all codewords, which is determined by $r$. 
As a comparison, the conventional uniform OIRS codebook with fixed parameters $\Delta \omega^u$ and $\Delta \gamma^u$ is chosen as a baseline scheme. 
Given an arbitrary PD location, the beam alignment is realized for both two OIRS codebooks and the channel gain errors between the selected beam and the real channel gain can be obtained. 
For fairness, the PD's location uniformly traverses all grid points on the detection plane with a spacing of 0.01m, such that the Frobenius norm of the error matrix can be calculated as the metric.

\begin{figure}[t]
	\centering
	\includegraphics[width=0.47\textwidth]{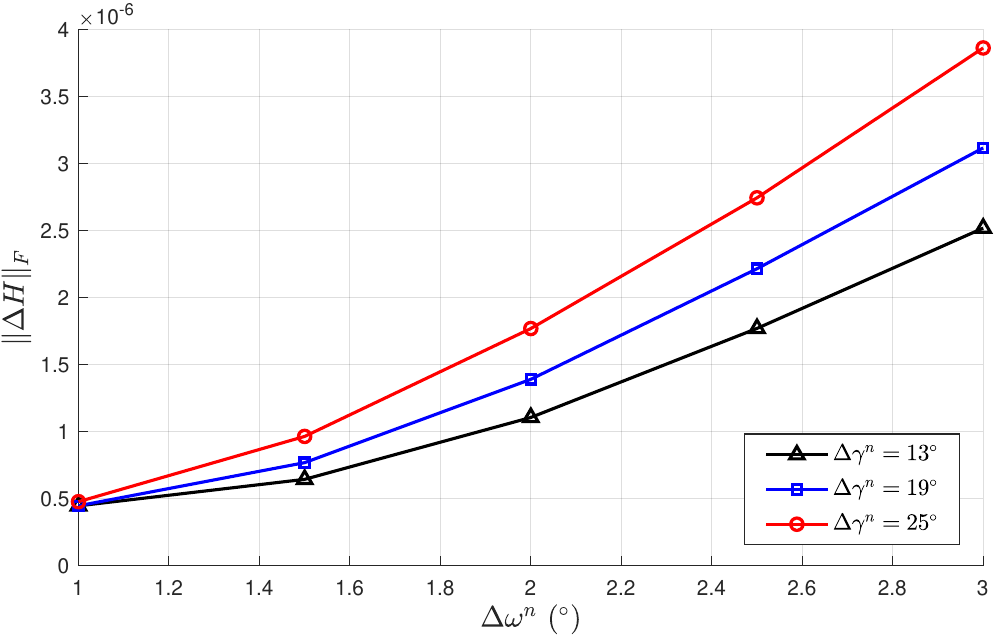}
	\caption{Frobenius norm of the channel error matrix versus $\Delta \omega^n$.}
	\label{Fig:omega}
\end{figure}

In Fig.~\ref{Fig:omega}, the performance of the proposed OIRS codebook is shown. 
Intuitively, $\omega^n$ determines the “radius” of each codeword while $\omega^n$ determines the interval length on the arc. 
It can be observed that the Frobenius norm decreases with the decline of $\Delta \omega^n$ due to the growing density of the codewords. 
When $\Delta \omega^n$ is sufficiently small, the proposed angle codeword has a high resolution such that keeping decreasing $\Delta \omega^n$ can barely improve the performance. 
Similarly, as shown in Fig.~\ref{Fig:gamma}, the Frobenius norm of the channel error matrix increases proportionally with $\Delta \gamma^n$, as denser angle codewords result in smaller error. 
However, the above results also reveal that the parameter $\Delta \omega^n$ plays a more significant role than the parameter $\Delta \gamma^n$. 
Particularly, the increase in $\Delta \omega^n$ is nearly tenfold compared to that in $\Delta \gamma^n$ under the same shift degree, which implies that the initialization of $\Delta \omega^n$ should be smaller than $\Delta \gamma^n$.

\begin{figure}[t]
	\centering
	\includegraphics[width=0.45\textwidth]{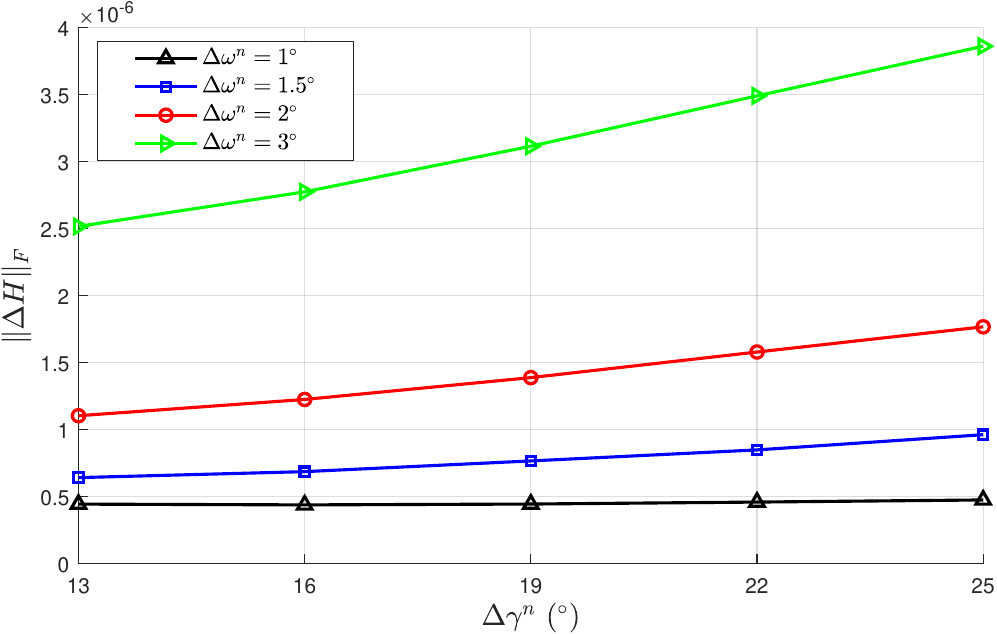}
	\caption{Frobenius norm of the channel error matrix versus $\Delta \gamma^n$.}
	\label{Fig:gamma}
\end{figure}

\begin{figure}[t]
	\centering
	\includegraphics[width=0.45\textwidth]{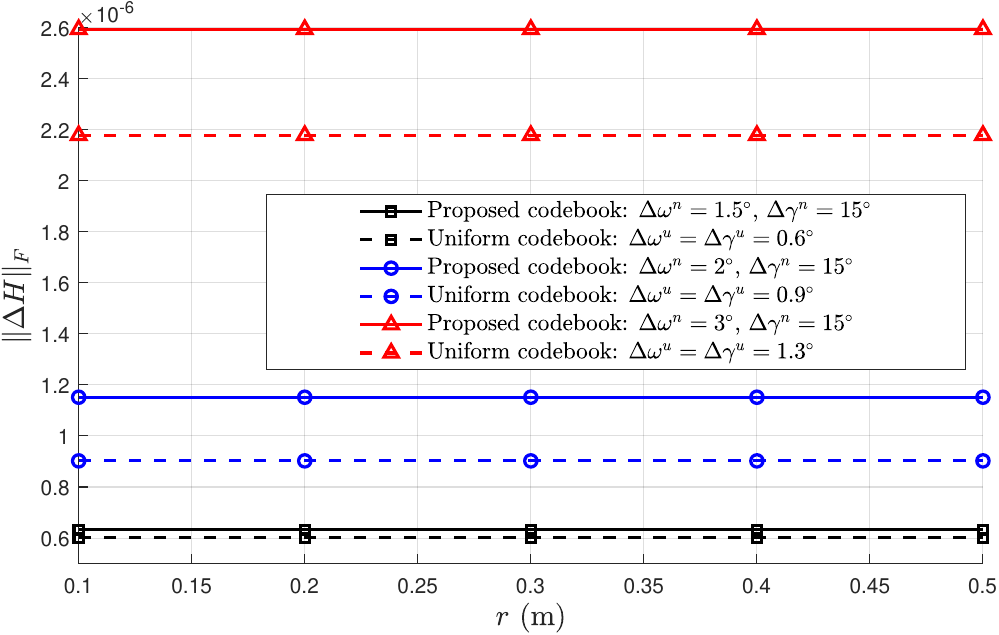}
	\caption{Frobenius norm of different OIRS codebook designs versus $r$.}
	\label{Fig:compare error}
\end{figure}

In Fig.~\ref{Fig:compare error}, the Frobenius norms of the uniform OIRS codebooks and the proposed GO-based non-uniform OIRS codebooks are shown. 
It can be observed that the performance metric of the uniform codebook, which is determined by parameters $\Delta \omega^u=\Delta \gamma^u=0.6\degree$, is nearly the same as the proposed codebook with the parameters $\Delta \omega^u=1.5\degree$ and $\Delta \gamma^u=15\degree$. 
Moreover, the uniform codebooks for $\Delta \omega^u=\Delta \gamma^u=0.9\degree$ and $\Delta \omega^u=\Delta \gamma^u=1.3\degree$ show comparable performance to the proposed codebook for $\Delta \omega^u=2\degree$ and $\Delta \gamma^u=15\degree$, and for $\Delta \omega^u=3\degree$ and $\Delta \gamma^u=15\degree$, respectively. 
As such, given that the real channel gain is $1.55\times 10^{-5}$ in average, the proposed OIRS codebook can achieve a similar beam alignment performance compared to the baseline, with less than $3$\% relative error. 
However, as depicted in Fig.~\ref{Fig:compare num}, the number of the candidate beams to be swept is significantly reduced in the proposed codebook. 
For example, considering $r = 0.5 \text{m}$, the searching number of the uniform codebook is $1,135$ when $\Delta \omega^u=\Delta \gamma^u=0.6\degree$, which is considerably higher than that of the proposed non-uniform codebook, i.e., $210$. 
The above results verify that the proposed OIRS codebook outperforms the baseline scheme and can improve the CSI acquisition of OIRS-assisted VLC.

\begin{figure}[t]
	\centering
	\includegraphics[width=0.45\textwidth]{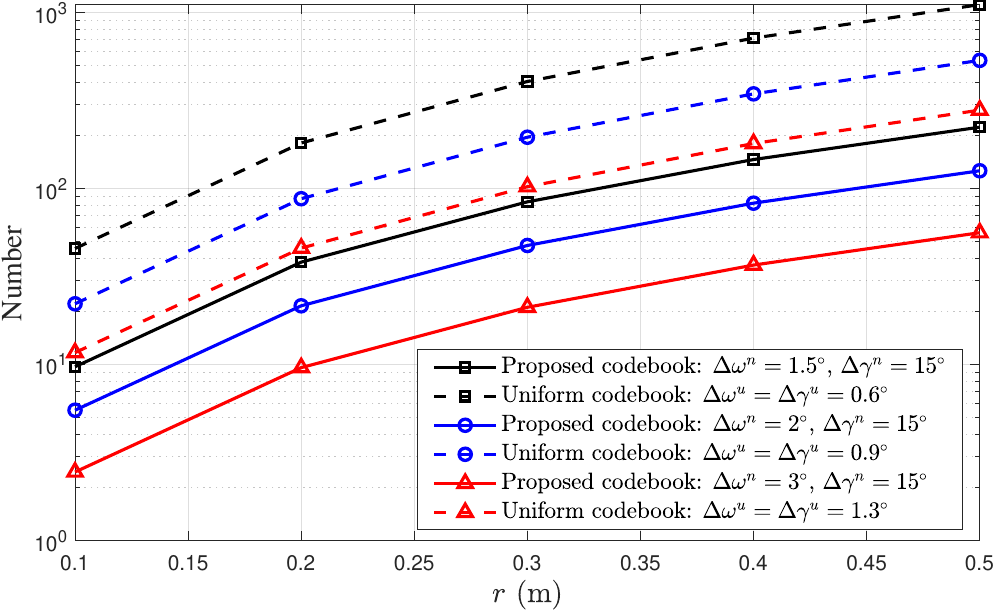}
	\caption{Comparison of the beam searching times between the proposed nonuniform codebook and the uniform codebook.}
	\label{Fig:compare num}
\end{figure}

\begin{figure}[t]
	\centering
	\includegraphics[width=0.45\textwidth]{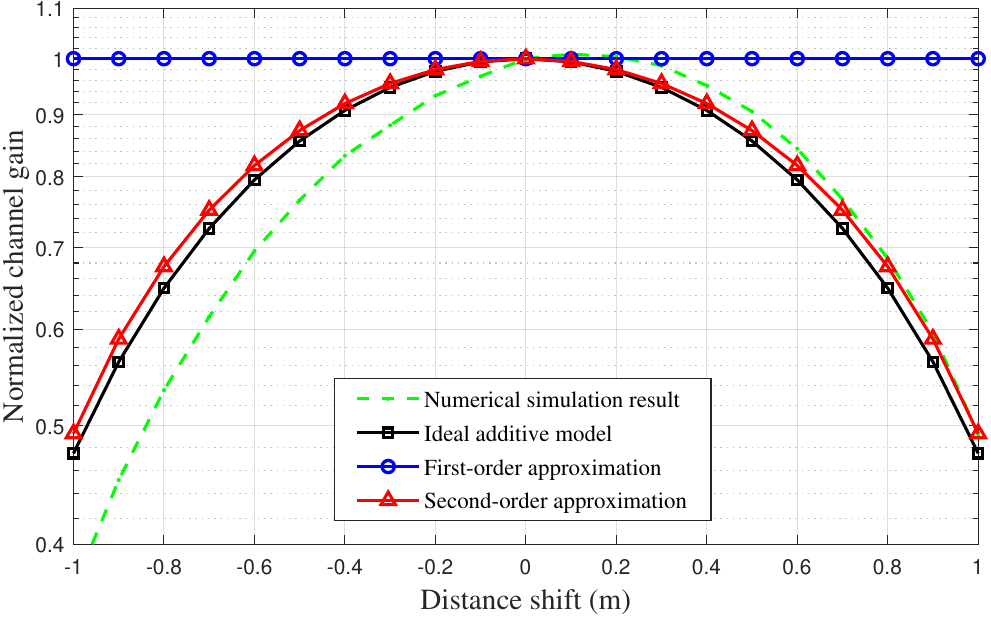}
	\caption{The spatial coherence among OIRS reflecting elements.}
	\label{Fig:distance}
\end{figure}

Next, to validate the spatial coherence of the OIRS, we simulate the normalized channel gain versus the distance shift when the location of the PD is set to (2m, 2m, 0m). 
The normal vector of the element is perpendicular to the wall such that the path of the LED-OIRS-PD adheres to the reflection law. 
As depicted in Fig.~\ref{Fig:distance}, the channel gain approximation obtained by the second-order Taylor's series in~(\ref{Eq:xis}) is more accurate compared to the linear approximation. 
Thus, when the coefficient $\xi_c$ is set to 0.04, it can be observed that the coherence distance of OIRS is nearly 0.4m, which can be employed to reduce the pilot overhead for the channel estimation. 
Moreover, the channel gain simulated by the physical optics is also illustrated. 
Although the ``additive'' Lambertian gain is not exactly accurate, the result shows that the OIRS coherence characteristic is still guaranteed.

\begin{figure}[t]
	\centering
	\includegraphics[width=0.45\textwidth]{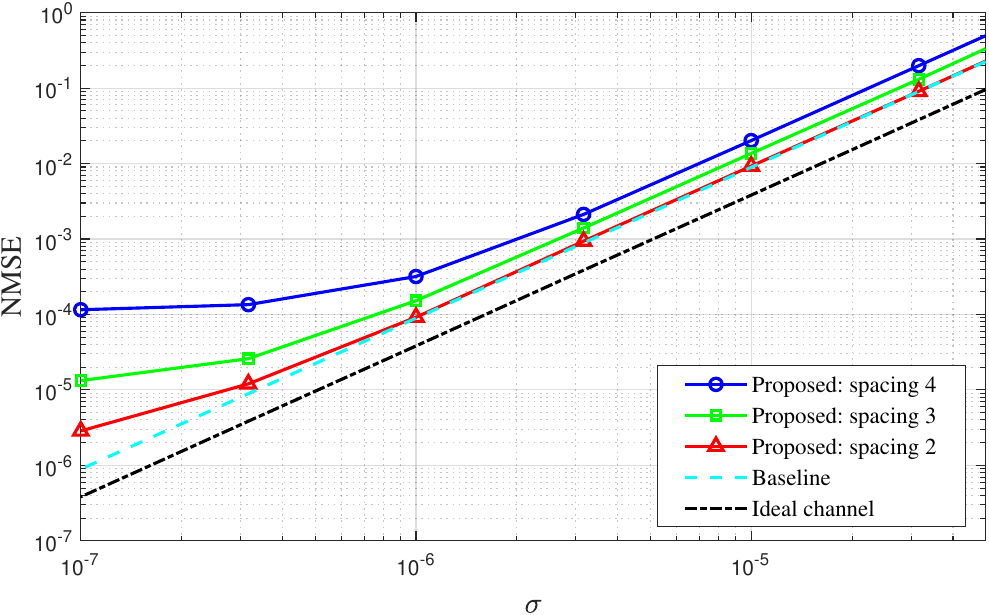}
	\caption{NMSE performance of the proposed OIRS channel estimation algorithm in the SISO case.}
	\label{Fig:NMSE_SISO}
\end{figure}

\begin{figure}[t]
	\centering
	\includegraphics[width=0.45\textwidth]{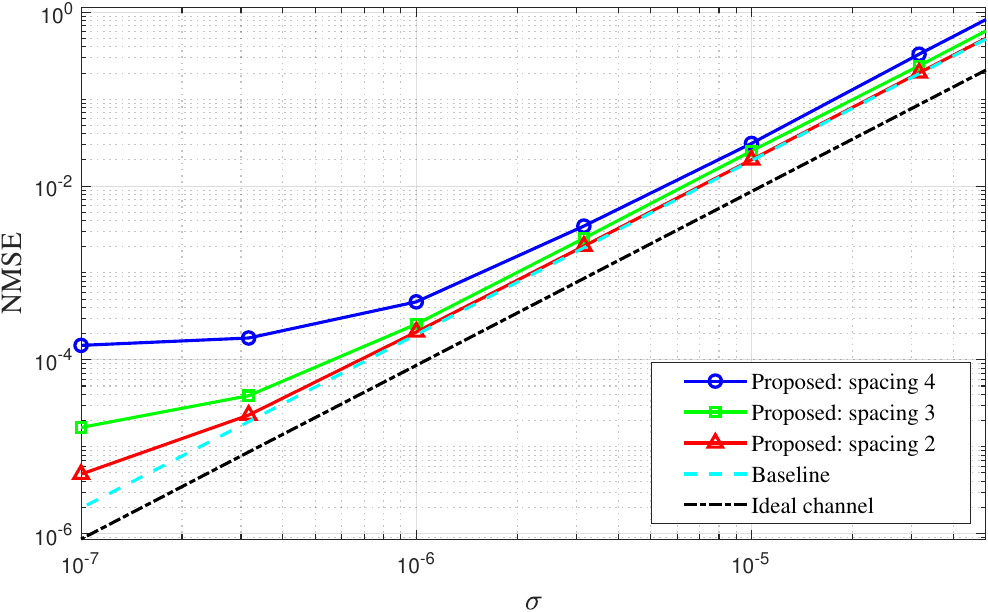}
	\caption{NMSE performance of the proposed OIRS channel estimation algorithm in the MIMO case.}
	\label{Fig:NMSE_MIMO}
\end{figure}

Afterwards, we evaluate the performance of the proposed JSTS algorithm. 
The OIRS with $N = 24^2 = 576$ elements is affixed to the $XoZ$ plane, and all reflecting elements are divided into several subarrays according to the coherence distance $d_c$. 
The spacing between any two subarrays can be obtained as $\lceil d_c/b \rceil$, which is equal to the square root of the number of OIRS elements in each subarray. 
Therefore, setting the spacing to $2$ means that the $576$ OIRS elements are divided into $12\times 12$ subarrays, where only the first element of each subarray is activated. 
Similarly, setting the spacing to $3$ and $4$ results in $8\times8$ subarrays and $6\times 6$ subarrays, respectively. 
For each activated subarray, $P = 100$ pilot symbols are generated to estimate the associated CSI following Algorithm~\ref{Algo:sampling}. 
In contrast, the baseline scheme that estimates $\tilde{\boldsymbol{H}}_c$ without spatial or temporal sampling is expected to achieve lower estimation error but higher CSI estimation overhead. 
As shown in Fig.~\ref{Fig:NMSE_SISO}, we plot the normalized mean square error (NMSE) of the OIRS channel estimation versus the standard deviation of the AWGN $\sigma$. 
It can be observed that the NMSE gap between the proposed algorithm and the baseline scheme is contingent on the sampling spacing. 
Specifically, the performance degradation is negligible when the spacing is less than the coherence distance, and the opposite holds true otherwise. 
Furthermore, given that the sampling method is derived from the Lambertian model in~(\ref{Eq:lambertian}), we also provide a comparison between the simulated channel gain and the ideal channel gain. 
The ideal scenario refers to the OIRS-reflected channel generated by the ``additive'' Lambertian model~(\ref{Eq:lambertian}), and the numerical result indicates that its NMSE is better than that in the practical scenario. 
This is consistent with the fact that the “additive” channel gain is an upper limit of the actual channel gain~\cite{abdelhady2020visible}.

\begin{figure}[t]
	\centering
	\includegraphics[width=0.45\textwidth]{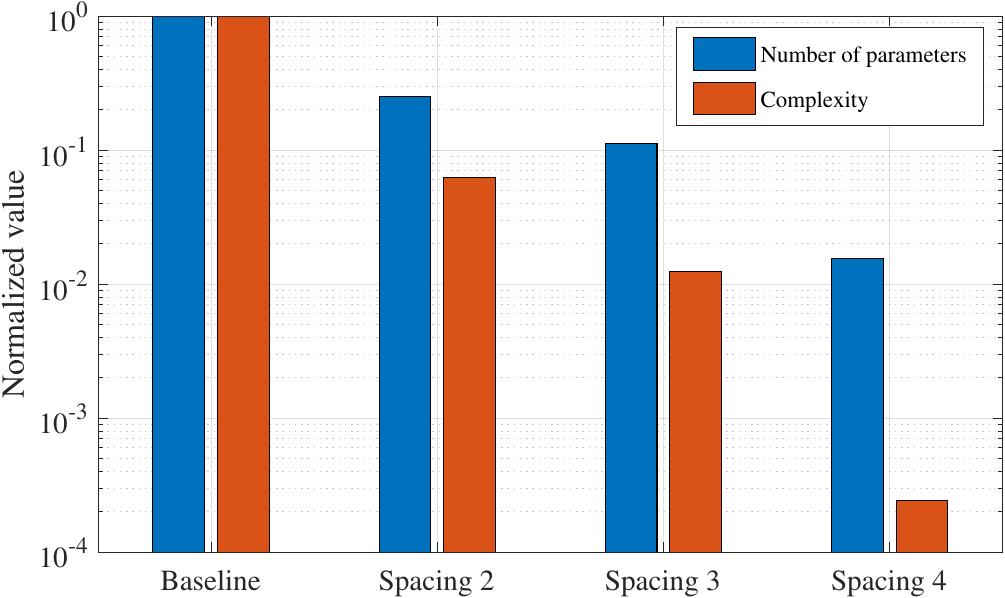}
	\caption{Overhead of the proposed channel estimation algorithm: Number of parameters and the computational complexity.}
	\label{Fig:overhead}
\end{figure}

Furthermore, we extend the simulation to the MIMO VLC case. 
The NMSE performance of the OIRS channel estimation is depicted in Fig.~\ref{Fig:NMSE_MIMO} with the number of LEDs and the number of PDs set to $N_t = N_r = 2$. 
For the scheme with the spacing equal to $2$, each element of a subarray is aligned with a specific pair of LED and PD. 
In this case, the spatial coherence of the OIRS-reflected channel is fully exploited and the NMSE gap is negligible. 
It is shown that the NMSE in the MIMO case is larger than that in the SISO case due to the channel interference. 
However, similar trend of NMSE curves can be observed from Fig.~\ref{Fig:NMSE_MIMO}, which implies that the proposed JSTS algorithm is effective for both SISO and MIMO VLC.

Finally, in Fig.~\ref{Fig:overhead}, we evaluate the overhead of the OIRS channel estimation. 
Due to the sampling method adopted in the proposed algorithm, the number of CSI parameters to be estimated decreases following the square law w.r.t. the spacing. 
For example, compared to the baseline scheme, our proposed algorithm with sampling spacing $2$ estimates the OIRS-reflected channel gains for every $4$ reflecting elements. 
Consequently, the total number of parameters for channel estimation is a quarter of that of the baseline. 
Next, the computational complexity of the MMSE estimator in~(\ref{Eq:MMSE estimator}) can be expressed as $\mathcal{O}(P(N_t^2 N_r + N_t^3 + N_t^2N_r)) = \mathcal{O}(PN_t^2(N_t + 2N_r))$.
Therefore, the complexity is inversely proportional to the square of the number of CSI parameters, which implies a significant complexity reduction of the proposed channel estimation algorithm. 
However, it can be observed that the NMSE of channel estimation increases when the sampling spacing becomes larger, which suggests a fundamental tradeoff between the channel estimation accuracy and overhead.

\section{Conclusions}
\label{Sec:Conclude}
This paper proposed a new channel estimation protocol for the OIRS-assisted VLC system. 
To reduce the exorbitant channel estimation overhead due to the high dimensional CSI matrix, we first derived the OIRS coherence distance and time for the OIRS-reflected channel. 
To achieve fast beam alignment over different channel coherence time, we proposed a GO-based non-uniform codebook by sampling the roll angle recursively and the yaw angle uniformly. 
Next, we leveraged the coherence distance of the OIRS and proposed a sequential channel estimation approach jointly with a space-time interpolation to acquire the full CSI within each channel coherence time Simulation results demonstrated the proposed OIRS codebook can achieve the same performance as the conventional uniform codebook with a considerably reduced searching space. 
In addition, our proposed samplingbased channel estimation can properly balance between the channel estimation accuracy and overhead, thus facilitating the practical implementation of the OIRS-assisted VLC systems.

\begin{appendices}
	\section{Proof of Lemma 1}
	\label{app:A}
	According to~(\ref{Eq:lambertian}), the OIRS-reflected channel gain with space shift $\Delta\textbf{R}$ is given by
	\begin{align}
		\label{Eq:h_spce_shift}
		h(\textbf{R} + \Delta\textbf{R}, \textbf{U}) \propto & \frac{1}{\left(\|\textbf{LR} + \Delta\textbf{R}\|_2 + \|\textbf{UR} + \Delta\textbf{R}\|_2\right)^2} \notag\\
		&\times \frac{\left(\widehat{\boldsymbol{N}}_1^T(\textbf{LR} + \Delta\textbf{R})\right)^m}{\|\textbf{LR} + \Delta\textbf{R}\|_2^m} 
		\frac{\widehat{\boldsymbol{N}}_2^T(\textbf{UR} + \Delta\textbf{R})}{\|\textbf{UR} + \Delta\textbf{R}\|_2}.
	\end{align}
	Based on $\nabla_{\boldsymbol{A}} \Psi(\boldsymbol{A})\Xi(\boldsymbol{A})\Gamma(\boldsymbol{A}) = \Xi(\boldsymbol{A})\Gamma(\boldsymbol{A})\nabla_{\boldsymbol{A}}\Psi(\boldsymbol{A}) + \Psi(\boldsymbol{A})\Gamma(\boldsymbol{A})\nabla_{\boldsymbol{A}}\Xi(\boldsymbol{A}) + \Psi(\boldsymbol{A})\Xi(\boldsymbol{A})\nabla_{\boldsymbol{A}}\Gamma(\boldsymbol{A})$, the derivative of the shifted channel gain in~(\ref{Eq:h_spce_shift}) is given by
	\begin{align}
		&\nabla_{\Delta\textbf{R}}h(\textbf{R} + \Delta\textbf{R}, \textbf{U}) \propto h(\textbf{R}, \textbf{U}) \Big\{ \frac{m}{d_1\cos\theta}\widehat{\boldsymbol{N}}_1 + \frac{1}{d_2\cos\phi}\widehat{\boldsymbol{N}}_2 \notag\\
		&- \left(\frac{m}{d_1} + \frac{2}{d_1 + d_2}\right)\widehat{\textbf{LR}}
		- \left(\frac{1}{d_2} + \frac{2}{d_1 + d_2}\right)\widehat{\textbf{UR}} \Big\},
	\end{align}
	following which the second-order derivative of $h(\textbf{R} + \Delta\textbf{R}, \textbf{U})$ can be obtained as
	\begin{align}
		\label{Eq:spa-Taylor second}
		&\nabla^2_{\Delta\textbf{R}}h(\textbf{R} + \Delta\textbf{R}, \textbf{U}) \notag\\
		\propto & h(\textbf{R}, \textbf{U})\Big\{\frac{-m}{d_1^2\cos^2\theta}\widehat{\boldsymbol{N}}_1\widehat{\boldsymbol{N}}_1^T - \frac{1}{d_2^2\cos^2\phi}\widehat{\boldsymbol{N}}_2\widehat{\boldsymbol{N}}_2^T \notag\\
		&+ 2\left(\frac{m}{d_1^2} + \frac{1}{(d_1 + d_2)^2} + \frac{2}{d_1^2(d_1 + d_2)}\right)\widehat{\textbf{LR}}\widehat{\textbf{LR}}^T \notag\\
		&+ 2\left(\frac{1}{d_2^2} + \frac{1}{(d_1 + d_2)^2}+\frac{2}{d_2^2(d_1 + d_2)}\right)\widehat{\textbf{UR}}\widehat{\textbf{UR}}^T \notag\\
		&+ \frac{2}{(d_1 + d_2)^2}\left(\widehat{\textbf{LR}}\widehat{\textbf{UR}}^T + \widehat{\textbf{UR}}\widehat{\textbf{LR}}^T\right) \notag\\
		&-\left(\frac{m}{d_1^2} + \frac{1}{d_2^2} + \frac{2}{d_1^2(d_1 + d_2)} + \frac{2}{d_2^2(d_1 + d_2)}\right)\boldsymbol{I}_3 \Big\}.
	\end{align}
	Therefore, the OIRS-reflected channel gain can be approximated according to the Taylor series expansion as
	\begin{align}
		\label{Eq:xis}
		h(\textbf{R} + & \Delta\textbf{R}, \textbf{U}) = h(\textbf{R}, \textbf{U}) + \nabla_{\Delta\textbf{R}}h(\textbf{R} + \Delta\textbf{R}, \textbf{U})^T\Delta\textbf{R} \notag \\
		& + \frac{1}{2}\Delta\textbf{R}^T\nabla^2_{\Delta\textbf{R}}h(\textbf{R} + \Delta\textbf{R}, \textbf{U})\Delta\textbf{R} + o\left(\|\Delta\textbf{R}\|^2\right),
	\end{align}
	and Lemma~\ref{Lemma:growth_rate_space} can be proven by substituting~(\ref{Eq:xis}) in~(\ref{Eq:xi_s_definition}).

	\section{Proof of Lemma 2}
	\label{app:B}
	According to~(\ref{Eq:lambertian}), the OIRS-reflected channel gain with time delay $\Delta t$ is given by
	\begin{align}
		\label{Eq:h_spce_time}
		h(\textbf{R}, \textbf{U}& + \Delta\textbf{U}) \notag\\
		\propto & \frac{\left(\widehat{\boldsymbol{N}}_1^T\widehat{\textbf{LR}}\right)^m}{\left(\|\textbf{LR}\|_2 + \|\textbf{UR} - \Delta\textbf{U}\|_2\right)^2}  \frac{\widehat{\boldsymbol{N}}_2^T(\textbf{UR} - \Delta\textbf{U})}{\|\textbf{UR} - \Delta\textbf{U}\|_2}.
	\end{align}
	Its first-order derivative is given by
	\begin{align}
		\nabla_{\Delta t}h(\textbf{R}, \textbf{U}+\boldsymbol{v}\Delta t)\propto 
		&h(\textbf{R}, \textbf{U}) \Big\{-\left(\frac{1}{d_2}+\frac{2}{d_1+d_2}\right)\widehat{\textbf{UR}}+\notag\\
		&\frac{1}{d_2\cos\phi}\widehat{\boldsymbol{N}}_2 \Big\}^T \boldsymbol{v},
	\end{align}
	and the second-order derivative of $h(\textbf{R}, \textbf{U}+\boldsymbol{v}\Delta t)$ can be expressed as
	\begin{align}
		&\quad \nabla^2_{\Delta t}h(\textbf{R}, \textbf{U}+\boldsymbol{v}\Delta t) \notag\\
		&\propto h(\textbf{R}, \textbf{U}) \boldsymbol{v}^T \Big\{ \frac{1}{d_2^2\cos^2\phi}\widehat{\boldsymbol{N}}_2\widehat{\boldsymbol{N}}_2^T -\left(\frac{1}{d_2^2} + \frac{2}{d_2^2(d_1 + d_2)}\right)\boldsymbol{I}_3 \notag\\
		&+ 2\left(\frac{1}{d_2^2} + \frac{1}{(d_1 + d_2)^2}+\frac{2}{d_2^2(d_1 + d_2)}\right)\widehat{\textbf{UR}}\widehat{\textbf{UR}}^T\Big\} \boldsymbol{v}.
	\end{align}
	Finally, based on the Taylor series, the OIRS-reflected channel gain can be approximated as
	\begin{align}
		\label{Eq:xit}
		h(\textbf{R}&, \textbf{U}+\boldsymbol{v}\Delta t) = h(\textbf{R}, \textbf{U}) + \nabla_{\Delta t}h(\textbf{R}, \textbf{U}+\boldsymbol{v}\Delta t)\Delta t \notag \\
		& + \frac{1}{2}\nabla^2_{\Delta t}h(\textbf{R}, \textbf{U}+\boldsymbol{v}\Delta )(\Delta t)^2 + o\left((\Delta t)^2\right),
	\end{align}
	and Lemma~\ref{Lemma:growth_rate_time} can be proven by substituting~(\ref{Eq:xit}) in~(\ref{Eq:xi_t_definition}).

	\section{Proof of Proposition 1}
	\label{app:C}
	The roots of the quadratic function in ~(\ref{Eq:xi_t}) can be obtained as $\Delta t_1= 0$ and~(\ref{Eq:delta_t2}) by solving the equation $\xi_t(\Delta t) = 0$. 
	According to the properties of quadratic functions, the derivative equals to $0$ when $\Delta t = \Delta t_2/2$, and the function value equals
	\begin{equation}
		\xi_t\left(\frac{\Delta t_2}{2}\right) = \frac{\Delta t_2}{4}\left\{\frac{1}{d_2\cos\phi}\widehat{\boldsymbol{N}}_2 -\left(\frac{1}{d_2}+\frac{2}{d_1+d_2}\right)\widehat{\textbf{UR}}\right\}^T\boldsymbol{v}.
	\end{equation}\\
	Regardless of the sign of $\Delta t_2$ and the convexity of the function $\xi_t(\Delta t)$, the solution to~(\ref{Eq:t_c_problem}) only depends on the relationship between $|\xi_t(\Delta t_2/2)|$ and $\xi_c$ due to the absolute operator in~(\ref{Eq:t_c_problem_constraint}). 
	Specifically, when $|\xi_t(\Delta t_2/2)|\leq\xi_c$, there will exist two roots for the equation $|\xi_t(\Delta t)|=\xi_c$. 
	According to Weda's Theorem, namely $(\Delta t_2 - \Delta t_1)^2 = (\Delta t_2 + \Delta t_1)^2 - 4\Delta t_1\Delta t_2$, the OIRS coherence time can be obtained by
	\begin{equation}
		t_c^2 = \left(|\Delta t_2|^2 + \frac{4\xi_c |\Delta t_2|}{\left|\Big\{\frac{1}{d_2\cos\phi}\widehat{\boldsymbol{N}}_2 - \left(\frac{1}{d_2} + \frac{2}{d_1 + d_2}\right)\widehat{\textbf{UR}} \Big\}^T \boldsymbol{v} \right|} \right)^2,
	\end{equation}
	which is an upper bound of $|\xi_t(\Delta t_2)|$. 
	Otherwise, there will be four roots when $|\xi_t(\Delta t_2/2)|\geq\xi_c$, which are intractable to obtain a precise formula. 
	Nonetheless, the OIRS coherence time can be upper-bounded by
	\begin{align}
		t_c =& 2 \times \frac{\Delta t}{2} \geq 2 \times \frac{\xi_c}{\nabla \xi_t|_{\Delta t = 0}} \notag\\
		=& \frac{2\xi_c}{\Big\{\frac{1}{d_2\cos\phi}\widehat{\boldsymbol{N}}_2 - \left(\frac{1}{d_2} + \frac{2}{d_1 + d_2}\right)\widehat{\textbf{UR}} \Big\}^T \boldsymbol{v}},
	\end{align}
	due to the decreasing derivative of the function $\xi_t(\Delta t)$.

	\section{Proof of Proposition 2}
	\label{app:D}
	Let $\Delta\textbf{R} \triangleq \Delta r \widehat{\Delta\textbf{R}}$, where $\widehat{\Delta\textbf{R}}$ and $\Delta r$ denote a specific direction and the amplitude of the space shift vector $\Delta\textbf{R}$, respectively. 
	It can be observed that $\xi_s(\Delta r \widehat{\Delta\textbf{R}})$ in~(\ref{Eq:xi_s}) has the same structure as $\xi_t(\Delta t)$ in~(\ref{Eq:xi_t}), and the roots of the equation $\xi_s(\Delta r)=0$ are $\Delta r_1=0$ and~(\ref{Eq:delta_r2}). 
	According to Theorem~\ref{Theorem:coherence_distance}, the coherence distance in the direction of $\widehat{\Delta\textbf{R}}$ can be obtained based on~(\ref{Eq:d_c}). 
	Therefore, the OIRS coherence distance $d_c$ is the miminum of $d_c(\widehat{\Delta\textbf{R}})$ over all possible directions.
\end{appendices}
\bibliographystyle{abbrv}
\bibliography{IEEEabrv,reference}
\end{document}